\documentclass{aa}
\usepackage{txfonts}
\usepackage{natbib}
\usepackage{graphicx,amssymb}
\usepackage{color}
\usepackage{longtable}
\usepackage{multirow}

\newcommand{\wave}[1]{$\lambda#1\,\mathrm{cm}$}  

\bibpunct{(}{)}{;}{a}{}{,} 
\def\alphan{\alpha_{\rm n}}

\def\degr{\hbox{$^\circ$}}
\def\farcm{\hbox{$.\mkern-4mu^\prime$}}

\def\radm{\,\mathrm{rad\,m^{-2}}}
\begin{document}

   \title{Magnetic fields and cosmic rays in M~31
   \thanks{Based on observations with the 100--m telescope of the Max-Planck-Institut f\"ur Radioastronomie at Effelsberg
   and with the VLA telescope of the National Radio Astronomy Observatory.}}
   \subtitle{II. Strength and distribution of the magnetic field components}

\titlerunning{Magnetic fields in M~31 II.}
\authorrunning{R.~Beck et al.}

   \author{R. Beck
          \and
          E.M. Berkhuijsen
          }

   \institute{Max-Planck-Institut f\"ur Radioastronomie, Auf dem H\"ugel 69, 53121 Bonn, Germany\\
              \email{rbeck@mpifr-bonn.mpg.de}
             }

   \date{Received 8 April 2025; accepted 1 July 2025}


  \abstract
      {Interstellar magnetic fields play an important role in the dynamics and evolution of galaxies.
      The nearby spiral galaxy M~31 is an ideal laboratory for extensive studies of magnetic fields.
      }
      {We measure the strength and distribution separately for the various magnetic field components:
      total, ordered, regular, isotropic turbulent, and, for the first time, anisotropic turbulent.
      }
      {Based on radio continuum observations of M~31 at 3.6\,cm and 6.2\,cm wavelengths with the
      Effelsberg 100--m telescope, plus combined observations with the VLA and Effelsberg
      telescopes at 20.5\,cm, the intensities of total, linearly polarized, and unpolarized synchrotron emission
      are measures of the strengths of total, ordered, and isotropic turbulent fields in the sky plane.
      We used two assumptions about equipartition between the energy densities of total
      magnetic fields and total cosmic rays, i.e. local equipartition and overall equipartition
      on the scale of order 10\,kpc and more. Faraday rotation measures (RMs) provided a model of the regular field.
      The quadratic difference between ordered and regular field strengths yields the strength
      of the anisotropic turbulent field.
      }
      {The average equipartition strengths of the magnetic field in the emission torus, between 8\,kpc and 12\,kpc radius
      in the galaxy plane,
      are $(6.3\pm0.2)\,\mu$G for the total, $(5.4\pm0.2)\,\mu$G for the isotropic turbulent, and
      $(3.2\pm0.3)\,\mu$G for the ordered field in the sky plane.
      The total, isotropic turbulent, and ordered field strength
      decrease exponentially with radial scale lengths of $\simeq$14--15\,kpc. The average strength
      of the axisymmetric regular field, $B_\mathrm{reg}$, derived from the RMs in the emission torus,
      is $(2.0\pm0.5)\,\mu$G and remains almost constant between 7\,kpc and 12\,kpc
      radius. Quadratic subtraction of the component $B_\mathrm{reg,\perp}$ in the sky plane from the
      ordered field, $B_\mathrm{ord,\perp}$, yields the strength of the anisotropic turbulent field,
      $B_\mathrm{an,\perp}$, which is $(2.7\pm0.7)\,\mu$G on average in the emission torus.
      Our test with an extreme non-equipartition case assuming constant CREs along the torus
      enhances the magnetic field fluctuations.
      }
      {The average strength of the regular field between 7\,kpc and 12\,kpc
      radius is about 40\% smaller than the equipartition
      strength of the ordered field (containing regular and anisotropic turbulent fields).
      As those two quantities were measured with independent methods,
      our results are consistent with the assumption of equipartition.
      Furthermore, our estimate of the diffusion length of cosmic-ray electrons (CREs) emitting
      at \wave{3.6} of $\lesssim 0.34$\,kpc in the sky plane sets the lower limit for the
      validity of the equipartition assumption.

      The average magnetic energy density in the emission torus is about five times
      larger than the thermal energy density of the diffuse warm ionized gas,
      while the magnetic energy density is similar to the kinetic
      energy density of turbulent motions of the neutral gas.
      Magnetic fields are a primary dynamical agent in the interstellar medium of M~31.

      The ratio between regular and isotropic turbulent fields is a measure of the relative efficiencies
      of the large-scale and the small-scale dynamos. The average value of $\simeq0.39$, almost
      constant with azimuth in the emission torus as well with radius in the range
      7--12\,kpc, is consistent with present-day dynamo models.
      The ratio between anisotropic and isotropic turbulent fields is $\simeq0.57$ on average and is almost
      constant with the azimuth in the emission torus as well as with the radius in the range
      7--10\,kpc. This indicates that anisotropic turbulent fields are generated by the shearing of isotropic
      turbulent fields.
      }


   \maketitle

   \nolinenumbers

\section{Introduction}
\label{sec:intro}

Interstellar magnetic fields play an important role in the structure
and evolution of galaxies. They provide support to the gas against the
gravitational field \citep{boulares90}, affect the star-formation rate
\citep[SFR;][]{taba18,krumholz19} and the multiphase structure of the interstellar medium
\citep[ISM;][]{ponnada22,gent24}, regulate galactic outflows and winds \citep{evirgen19}, and
control the propagation of cosmic rays \citep[e.g.][]{zweibel13}.

Synchrotron radio emission is the best tool to study magnetic fields in the ISM. Synchrotron intensity depends on the strength of the field
components in the sky plane and the number density of cosmic-ray electrons (CREs) in the
relevant energy range. To measure the field strength \text{B}, additional information about the
CRE number density or a relation between B and the CRE number density is needed.
Unpolarized synchrotron emission traces magnetic fields that cannot be resolved
by the telescope beam, i.e. isotropic turbulent fields
or ordered fields with orientations varying on scales smaller than the beam size.
Linearly polarized synchrotron emission is a signature of the strength and orientation of
resolved ordered fields in the sky plane;
these can be regular fields or anisotropic turbulent fields.

Regular (`mean') fields, $B_\mathrm{reg}$, are generated by the mean-field $\alpha$--$\Omega$ dynamo
(e.g. \citet{beck96,chamandy16}, \citet{brandenburg23}, Chapter~7 in \citet{shukurov21}), reveal a coherent
direction over several kiloparsecs.
The mean-field dynamo generates azimuthal magnetic field modes that can be
identified via large-scale variations in Faraday rotation measures \citep[RMs; Table~5 in][]{beck19}.

Isotropic turbulent fields, $B_\mathrm{turb}$, are generated by turbulent gas motions,
the small-scale dynamo (e.g. \citet{brandenburg05} and Chapter~6 in \citet{shukurov21}),
and can reach an energy density of some fraction of the energy density of turbulence.
Tangling of the regular magnetic field by turbulence also produces isotropic
turbulent fields with an energy density similar to that of the regular field (Appendix~A
in \citet{seta18}, Section~4.1 in \citet{seta20}, and Section~13.3 in \citet{shukurov21}).

Anisotropic turbulent fields, $B_\mathrm{an}$, can be generated from turbulent
fields by shearing or compressing gas flows; for example, shear by differential rotation
\citep[Section~5 in][]{hollins17} with a larger dispersion in azimuthal orientation,
or compression by density-wave shocks with a larger dispersion along the shock front.
Anisotropic turbulent fields reverse their sign on the scales of turbulence.
\footnote{We define the anisotropic turbulent field as a field, in which the vectors
are perfectly parallel to each other on the scale of the telescope beam size but
reverse their directions at random.}
The degree of anisotropy, $\delta$, is computed as $\delta=(1+B_\mathrm{an}^2/B_\mathrm{turb}^2)^{\,0.5}$.

Both anisotropic turbulent and regular fields give rise to linearly polarized emission.
To distinguish these two components, additional observations of the regular field are required,
such as Faraday RM or Zeeman spectral line splitting.
The latter can be observed in dense clouds in the Milky Way but not yet in external galaxies.
Faraday rotation of the polarization angle of synchrotron emission increases with the square
of the wavelength, the density of thermal electrons, and the average strength of the
field components along the line of sight. The sign of RM gives the field direction.
As RM is sensitive only to the regular field component along the line of sight, its component
in the sky plane remains unknown. This fact hinders the construction of a 3D model of the
large-scale regular field in a galaxy.

The Andromeda galaxy M~31 (NGC~224) is the nearest spiral galaxy at a distance
of 780\,kpc,
inclined by $75^\circ$ with respect to the sky plane,
see Table~1 of \citet{beck20} for basic parameters.
Thanks to its proximity, M~31 is particularly suited for investigating the properties of the
ISM.
The total and linearly polarized emissions are concentrated in a toroidal structure
between radii of about 8\,kpc and 12\,kpc from the galaxy's centre. This is the region with the
highest density of cold molecular gas \citep{nieten06}, warm neutral gas
\citep{brinks84,braun09,chemin09}, warm ionized gas \citep{devereux94},
and dust \citep{gordon06,fritz12}, and it is the location of most of the
present-day star formation \citep[e.g.][]{taba10,rahmani16}.

The polarization surveys of M~31 at \wave{6.2} and \wave{11.1} observed with
the Effelsberg 100--m telescope enabled us to calculate RMs,
which were interpreted as a large-scale axisymmetric spiral (ASS) pattern
of the regular field, which is regarded as the lowest mode excited by the mean-field
($\alpha$--$\Omega$) dynamo \citep{berkhuijsen03}. Combined with another
polarization survey at \wave{20.5} observed with the Very
Large Array (VLA) D-array and the Effelsberg telescope \citep{beck98},
a model of the large-scale magnetic field was constructed by
\citet{fletcher04}.

With the advent of more sensitive receivers, three new surveys of M~31 were
conducted with the Effelsberg telescope at \wave{3.6}, \wave{6.2}, and \wave{11.3}
with significantly improved sensitivities and map extensions \citep{beck20}.
A survey of M~31 at \wave{4.5} (6.6\,GHz) covering a large field
of $2.4\degr \times 3.1\degr$ was conducted by \citet{fatigoni21} with
the 64--m Sardinia Radio Telescope (SRT).

This paper presents the data used for this work in Section~\ref{sec:data},
Section~\ref{sec:diff} discusses the diffusion of CREs.
Section~\ref{sec:results} presents the maps of magnetic field strengths.
Section~\ref{sec:radial} shows the radial variation in field strengths.
Section~\ref{sec:energy} compares the energy densities of the various
components of the ISM in M~31, and
Appendix~\ref{sec:dp} discusses the origin of Faraday depolarization.

\section{Data}
\label{sec:data}

Our investigations are based on the radio continuum maps at \wave{3.6} (8.35\,GHz)
and \wave{20.5} (1.465\,GHz) presented by \citet{beck20}.
The original resolutions are $83\arcsec$ and $45\arcsec$, respectively.
Background sources with flux densities above 5\,mJy at \wave{20.5} and above 1.2\,mJy
at \wave{3.6} were subtracted. The central source of M~31 was also subtracted.
Both maps were smoothed to a common resolution of $1\farcm5$.
Furthermore, the map at \wave{6.2} (4.85\,GHz) at $3\arcmin$ resolution by \citet{beck20}
was used for the regions where the \wave{3.6} map lacks large-scale emission.

\begin{figure}[htbp]
\centering
\includegraphics[width=0.95\columnwidth]{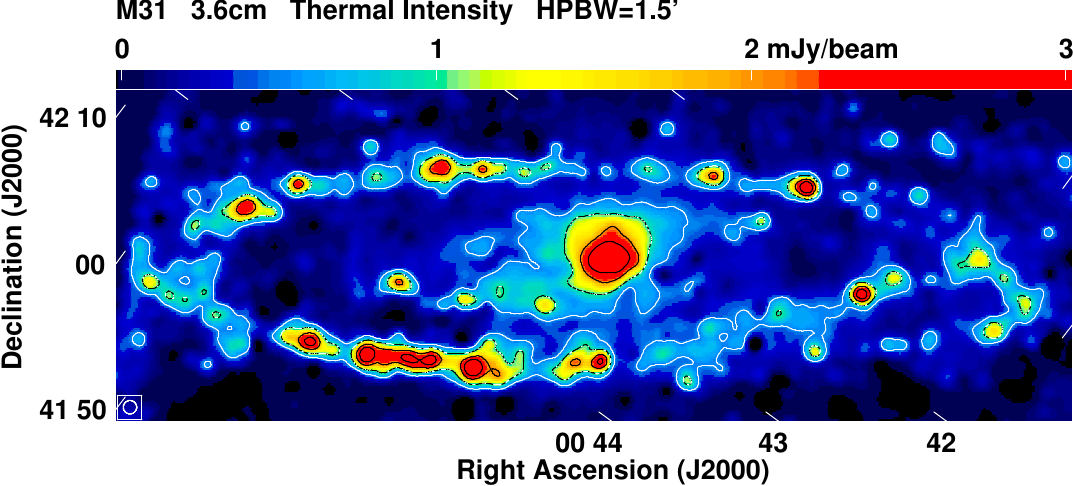}
\caption{Thermal intensity of M~31 at \wave{3.6} with $1\farcm5$ resolution.
The rms noise is $\simeq0.3$\,mJy/beam.
Contour levels are at 0.5, 1, 2, and 3\,mJy/beam.
Bright background sources have been subtracted.}
\label{fig:cm3th}
\centering
\end{figure}

\begin{figure}[htbp]
\centering
\includegraphics[width=0.95\columnwidth]{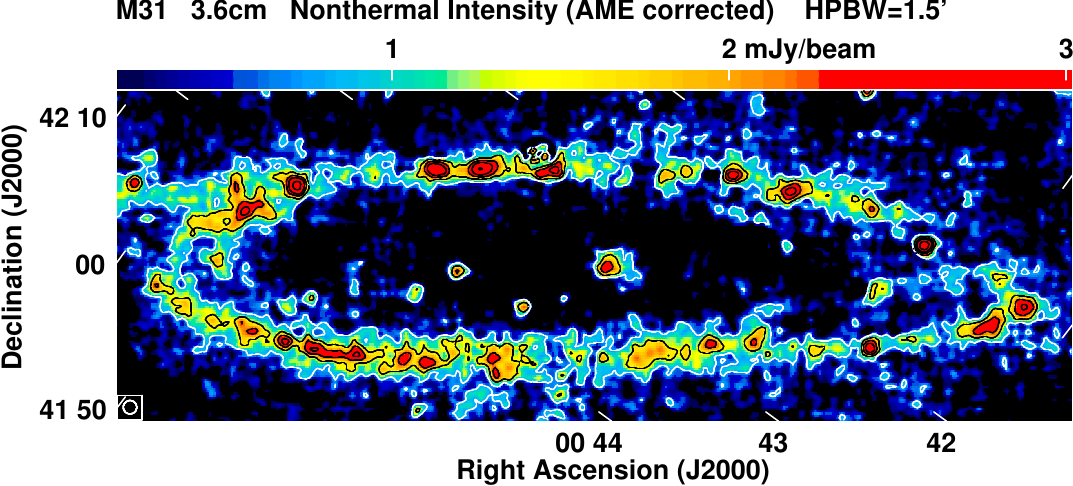}\\
\vspace{4mm}
\includegraphics[width=0.95\columnwidth]{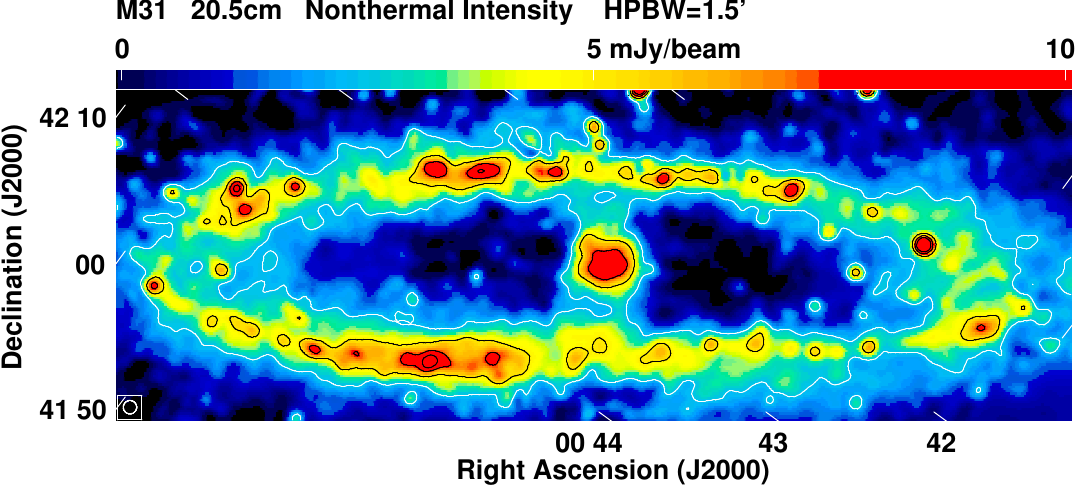}\\
\vspace{4mm}
\includegraphics[width=0.95\columnwidth]{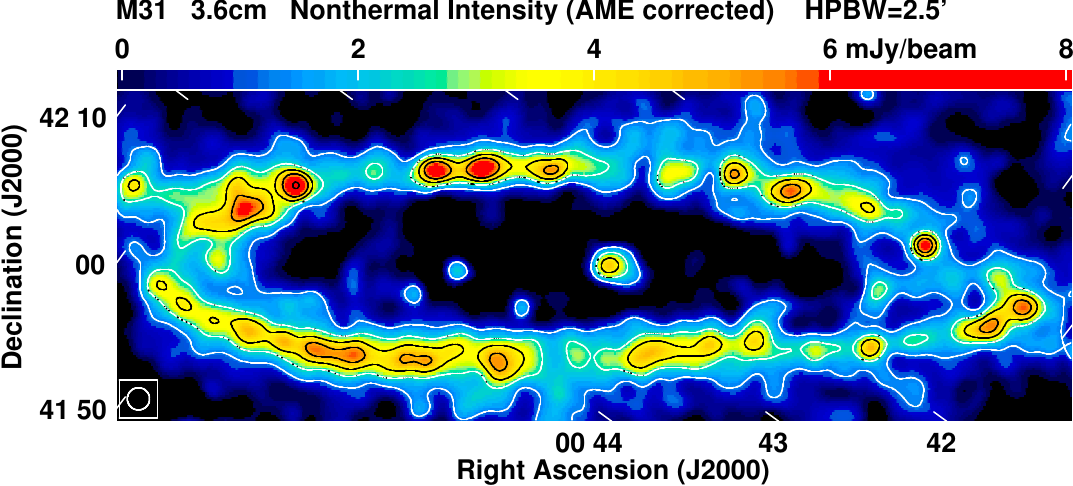}
\caption{Non-thermal intensity of M~31 at \wave{3.6} with $1\farcm5$ resolution (corrected for
AME) (\textbf{top}), at \wave{20.5} with $1\farcm5$ resolution (\textbf{middle}),
and at \wave{3.6} with $2\farcm5$ resolution (corrected for AME) (\textbf{bottom}).
The rms noise is $\simeq0.25$\,mJy/beam (top), $\simeq0.25$\,mJy/beam (middle), and
$\simeq0.4$\,mJy/beam (bottom). Contour levels are at 0.75, 1.5, 2.25, 3.0, and 4.5\,mJy/beam (top),
2.5, 5.0, 7.5, 10, 15, 20\,mJy/beam (middle), and 1.25, 2.5, 3.75, and 7.5\,mJy/beam (bottom).
The half-power beam widths are indicated in the bottom left corner.
Bright background sources and the unresolved central source of M~31 have been subtracted.
At \wave{3.6}, large-scale emission is missing in the central region.
}
\label{fig:cm3+20nth}
\centering
\end{figure}

The original map at \wave{20.5} revealed weak diffuse emission on scales larger than
the map size at \wave{3.6}. This was removed by a subtraction of an
overall base level of 0.3\,mJy/beam.

\begin{figure}[htbp]
\begin{center}
\includegraphics[width=0.95\columnwidth]{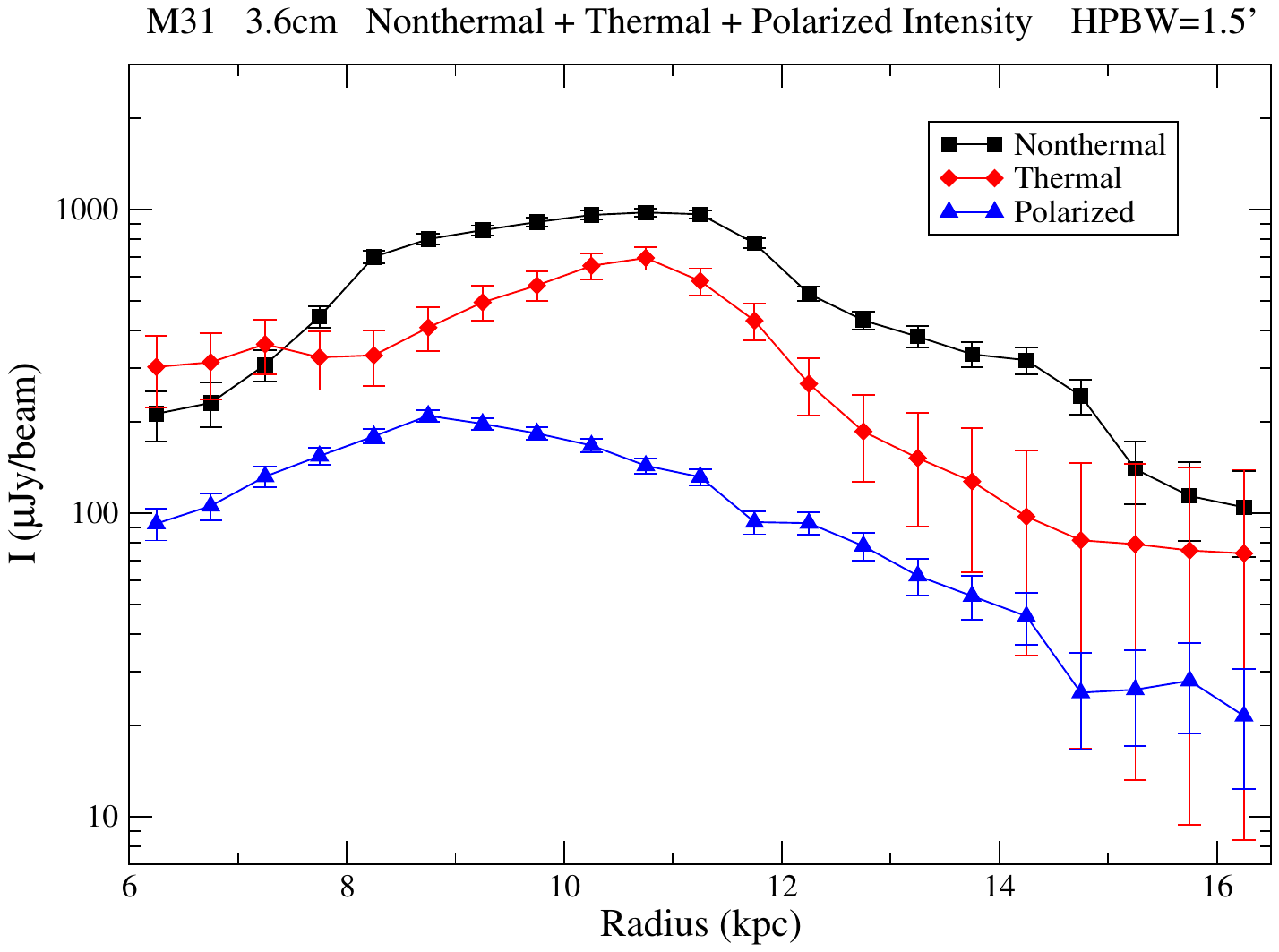}
\caption{Radial variation in the intensities (in log scale) of the non-thermal
(corrected for AME), thermal, and polarized emission at \wave{3.6} at
$1\farcm5$ resolution averaged in 0.5--kpc wide rings in the galaxy plane.
In this and the following figures, the error bars show the 1-$\sigma$ uncertainties.
}
\label{fig:radial_I}
\end{center}
\end{figure}

\subsection{Thermal--non-thermal separation}

In order to obtain maps of non-thermal (synchrotron) intensity (NTH) at these frequencies,
we subtracted maps of thermal emission (TH) from the maps of total emission (I).
We used the thermal map at \wave{20.5} that
\citet{taba13a} derived from the extinction-corrected H$\alpha$ map of
\citet{devereux94}, smoothed to HPBW = $1\farcm5$ and scaled to
\wave{3.6} by $(3.6/20.5)^{0.1}$, assuming a thermal spectral index
\footnote{We use the definition $I\propto\nu^{-\alpha}$ throughout this paper.} of $\alpha_{\rm th}=0.1$.
The resulting map is shown in Figure~\ref{fig:cm3th}.
The average thermal contribution to the total flux density in the radial range 8--12\,kpc
is 45\% at \wave{3.6} and 15\% at \wave{20.5}.

\begin{figure}[htbp]
\begin{center}
\includegraphics[width=0.95\columnwidth]{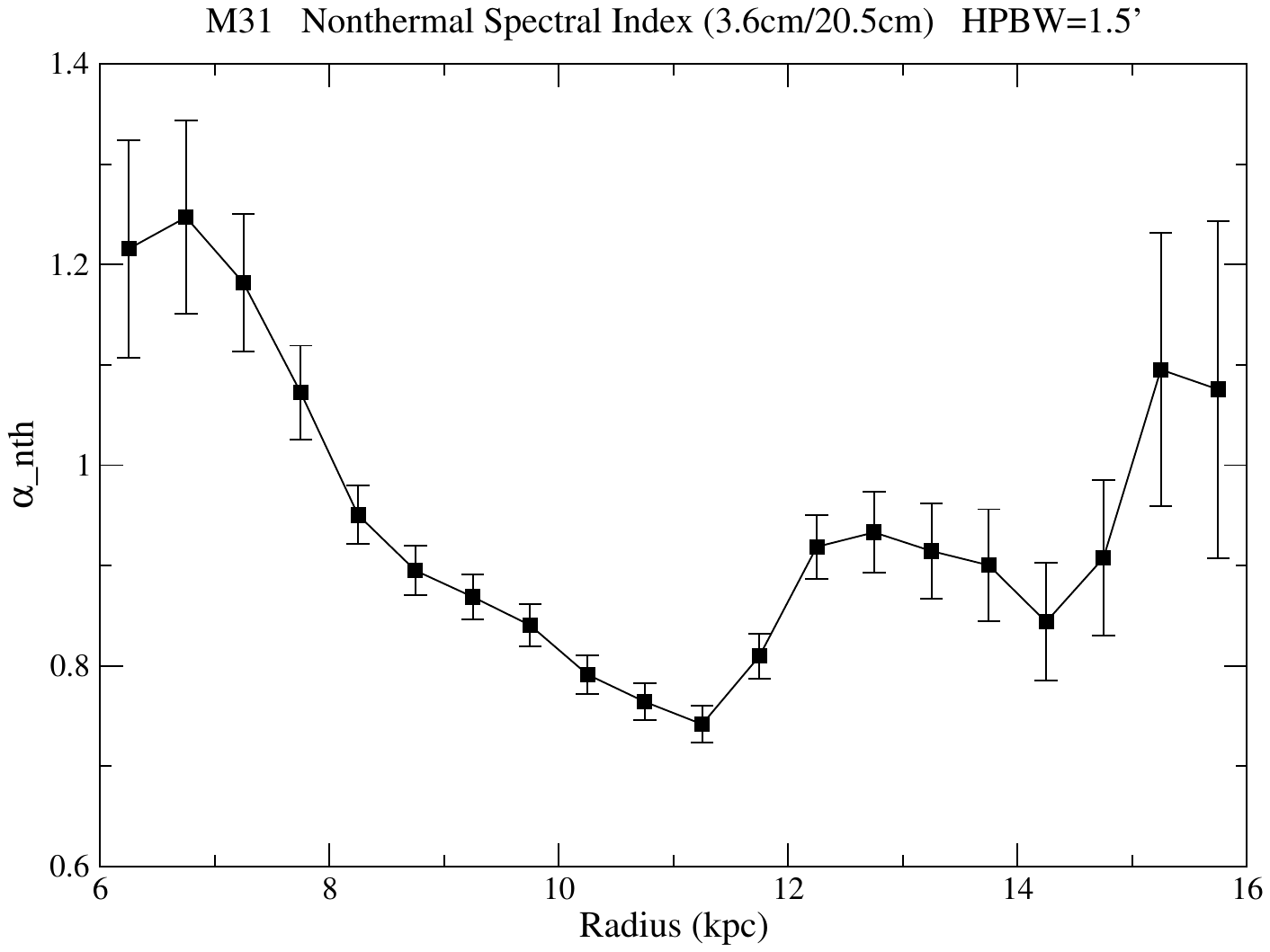}\\
\includegraphics[width=0.95\columnwidth]{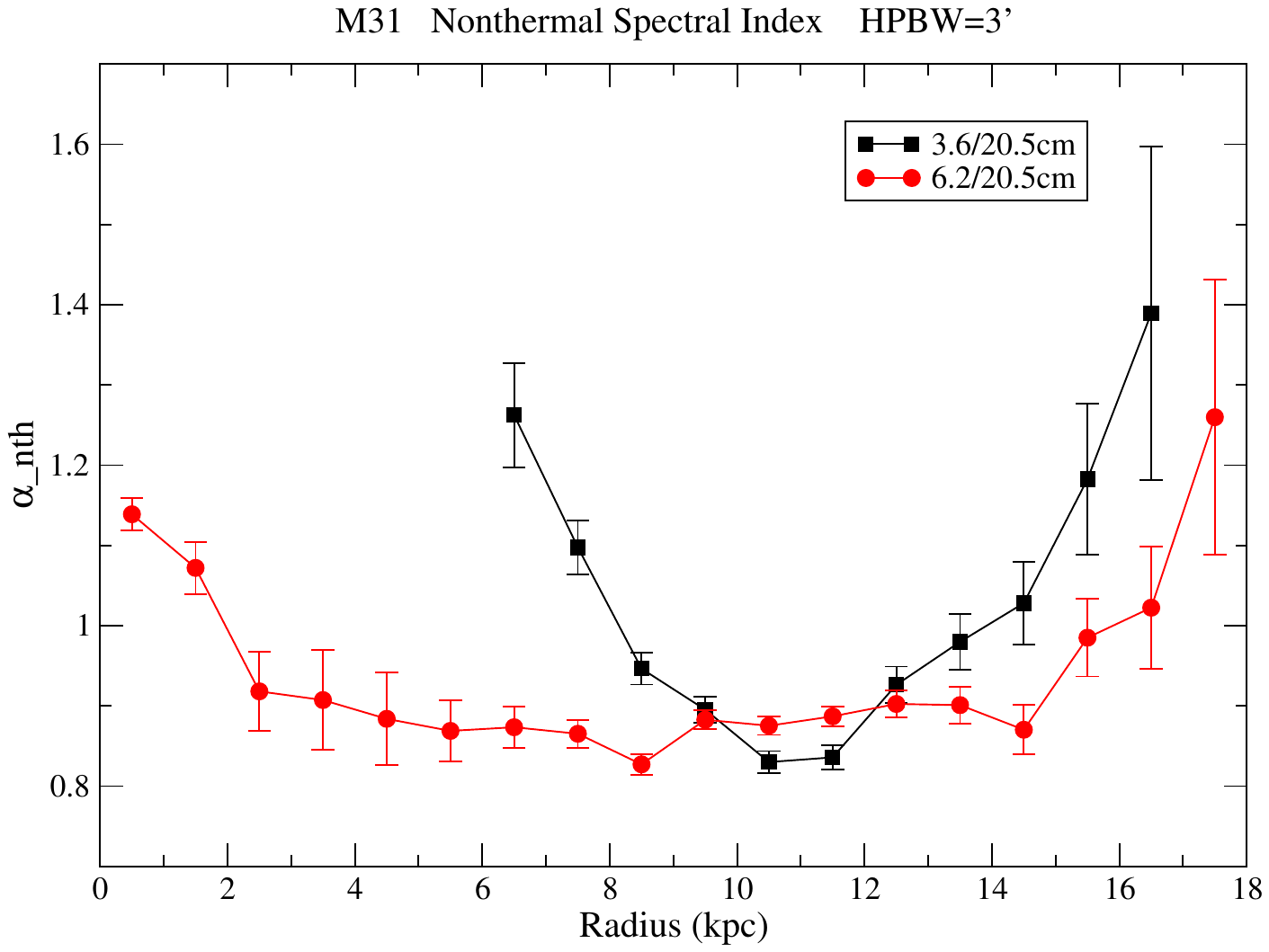}\\
\includegraphics[width=0.95\columnwidth]{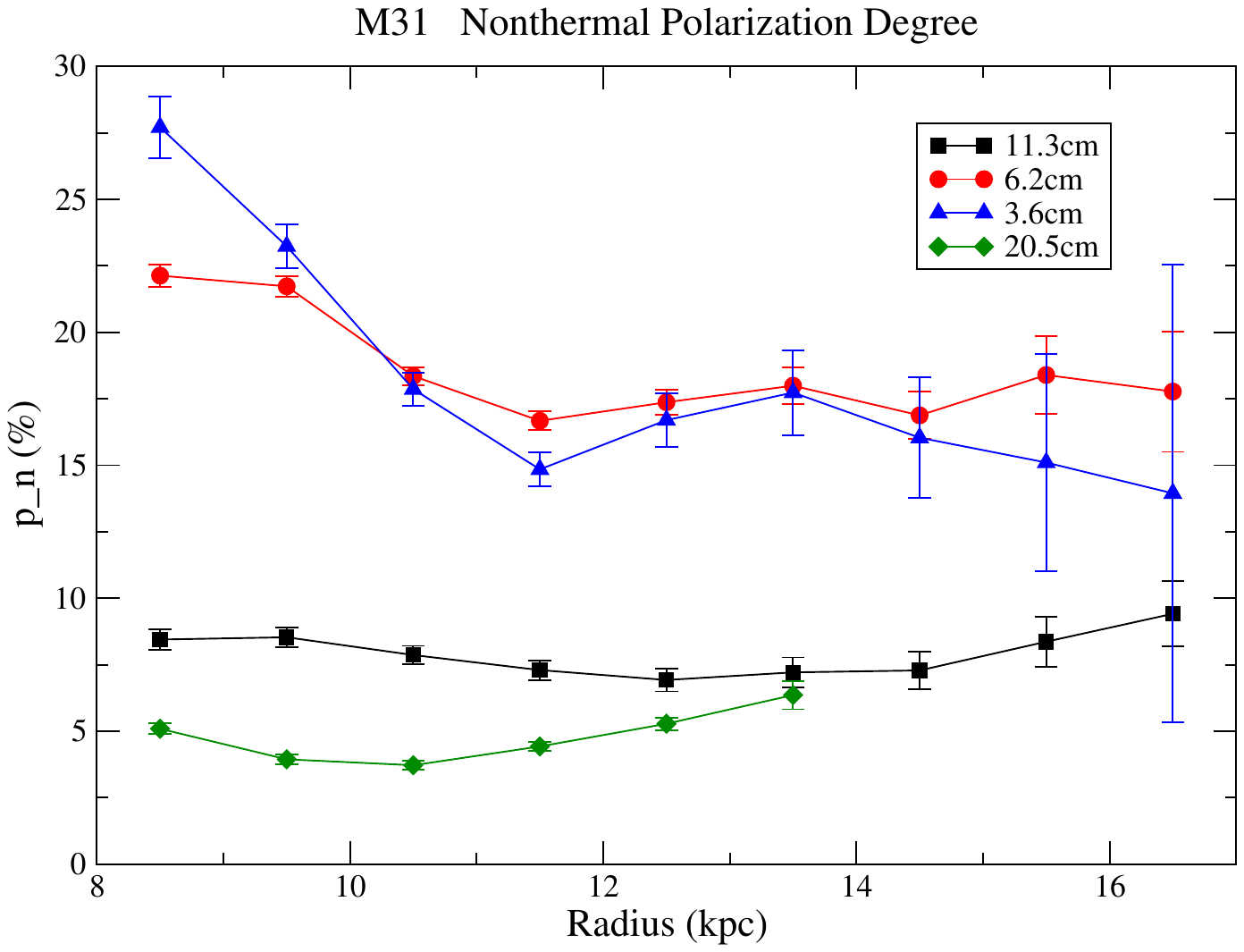}
\caption{\textbf{Top:} Radial variation in the average spectral index, $\alphan$, of the non-thermal
intensities  at $1\farcm5$ resolution between \wave{3.6} (corrected for AME) and
\wave{20.5} in 0.5--kpc wide rings in the galaxy plane.
\textbf{Middle:} Radial variations in the average spectral index, $\alphan$, of non-thermal
intensities at $3\arcmin$ resolution between \wave{3.6} (corrected for AME) and \wave{20.5}
as well as between \wave{6.2} and \wave{20.5} in 1--kpc wide rings in the galaxy plane.
\textbf{Bottom:} Radial variations in the average degrees of polarization, $p_\mathrm{n}$,
of the nonthermal intensities at \wave{3.6} (corrected for AME) and at \wave{6.2}, both at
$3\arcmin$ resolution, at \wave{11.3} at $5\arcmin$ resolution, and at \wave{20.5} at
$3\arcmin$ resolution.
}
\label{fig:radial}
\end{center}
\end{figure}

\subsection{Correction for anomalous microwave emission (AME)}
\label{sec:AME}

The total spectral indices between \wave{3.6} and \wave{20.5} (Figure~13 in \citet{beck20})
are smaller (flatter) than the total spectral indices between \wave{4.5} (6.6\,GHz) and \wave{20.5}
shown in Figure~12 of \citet{fatigoni21}. Furthermore, the non-thermal spectral indices in Figure~14
of \citet{beck20} are smaller (flatter) than the non-thermal spectral indices between
\wave{6.2} (4.85\,GHz) and \wave{20.5} \citep{berkhuijsen03}.

One reason for this discrepancy could be a contribution of anomalous microwave emission (AME)
by ultra-small dust grains (`spinning dust'),
as has been proposed by \citet{fatigoni21}. From the AME flux density spectra of M~31 measured with low
resolution by \citet{fernandez24}, referring to an integration area of $42\times27$\,kpc, the
flux density at 8.35\,GHz is $(0.44\pm0.37)$\,Jy, which is still an upper limit (Fernandez-Torreiro,
priv. comm.). The large uncertainty arises from the not perfectly matching apertures, the
different filtering techniques between instruments, different point source subtractions,
and a possible contamination from the Galactic foreground. We measured a total flux density
at \wave{3.6} of $(1.41\pm0.20)$\,Jy over an integration area of $16\times4$\,kpc \citep{beck20},
so that AME contributes less than 31\% of the total flux density at \wave{3.6}.

In a pragmatic approach, we assumed that the spectrum of the synchrotron flux density integrated over
the emission torus (between 8\,kpc and 12\,kpc radius) at $3\arcmin$ resolution is constant between
\wave{20.5} and \wave{3.6}. The non-thermal spectral index between \wave{20.5} and \wave{6.2}
is $0.87\pm0.01$. To achieve the same non-thermal spectral index between \wave{20.5} and \wave{3.6},
the intensities at \wave{3.6} have to be reduced by $16.9\pm0.5\%$. Assuming that the relative AME
contribution is constant across the galaxy, we applied a correction factor of 0.831 to all
non-thermal intensities at \wave{3.6}. The corrected map is presented in Figure~\ref{fig:cm3+20nth}
(top panel). As AME is not polarized, polarized intensities are not affected.

The correction removes the discrepancy between the Effelsberg data at \wave{3.6} and the SRT data
at \wave{4.5} \citep{fatigoni21} only partly. We cannot exclude that
the AME contribution is larger than estimated above. However, this should lead to a steepening in
the synchrotron spectrum between \wave{6.2} and \wave{3.6}. The corresponding break in the CRE
energy spectrum would need a synchrotron lifetime of CREs that is shorter than the residence time
in the torus, which is not the case (see Section~\ref{sec:diff}).

\subsection{Final non-thermal maps}
\label{sec:final}

The emission from the toroidal structure between about 8\,kpc and 12\,kpc from the galaxy's centre is
clumpier at \wave{3.6} than at \wave{20.5} (Fig.~\ref{fig:cm3+20nth}).
Strong emission emerges from regions with high SFRs evident from
their H$\alpha$ emission \citep{devereux94}, as discussed by \citet{taba10}.

The central region of M~31 appears weaker at \wave{3.6} than at \wave{20.5}. The map at \wave{3.6}
was combined from observations of seven small regions. Adjusting base levels when combining
these regions was difficult and tended to suppress the extended emission. As a result, the diffuse total
(and non-thermal) emission in the central region is weaker at \wave{3.6} compared to the maps
at longer wavelengths that were observed in one single field. Consequently, the non-thermal
intensities at \wave{3.6} within 8\,kpc radius are not used in the following.

Figure~\ref{fig:radial_I} shows the radial variations in the average thermal, non-thermal,
and polarized intensities at \wave{3.6} in the radial range 6--16.5\,kpc. The uncertainties
increase beyond 15\,kpc radius. Fitting exponential functions $I\propto \exp(-R/R_{3.6})$
in the radial range 11--15\,kpc gives scale lengths of $R_\mathrm{3.6,th}=(1.7\pm0.2)$\,kpc and
$R_\mathrm{3.6,nth}=(2.7\pm0.3)$\,kpc.
The non-thermal emission at \wave{20.5} with $3\arcmin$ resolution (not shown here) was detected
until 20\,kpc radius with small uncertainties. The scale length in the radial range 11--20\,kpc
is $R_\mathrm{20.5,nth}=(3.1\pm0.1)$\,kpc.

\subsection{Non-thermal (synchrotron) spectral index}

The AME correction applied to the intensities at \wave{3.6} increases (steepens) all non-thermal
spectral indices between \wave{3.6} and \wave{20.5} by a constant offset of 0.106 compared to
Fig.~14 of \citet{beck20}.
Figure~\ref{fig:radial} (top panel) shows the radial variation in the average spectral indices
between \wave{3.6} and \wave{20.5} between 6\,kpc and 16\,kpc radius.
The steepening of $\alphan$ beyond 15\,kpc radius is weaker than in Fig.~15 of \citet{beck20}.
The steep spectral index in the radial range 6--9\,kpc is an artefact of missing
diffuse \wave{3.6} emission in the central region (see above).

The average spectral indices between \wave{3.6} and \wave{20.5} and between \wave{6.2} and \wave{20.5}
at $3\arcmin$ resolution are compared in Figure~\ref{fig:radial} (middle panel). The agreement is
good between 9\,kpc and 13 \,kpc radius. The spectral steepening between \wave{3.6} and \wave{20.5}
at smaller and larger radii could be due to energy losses (synchrotron and/or Inverse Compton)
of CREs emitting at \wave{3.6} when propagating away from their places of origin in the torus
at 8--12\,kpc radius.
Another reason could be that some large-scale emission is missing at \wave{3.6} due to the low
signal-to-noise ratios outside the torus.
Total intensity data at \wave{3.6} outside 8--12\,kpc radius will not be included in the further
analysis of this paper.

\subsection{Polarization degree}

Figure~\ref{fig:radial} (bottom panel) shows the average radial variation in the polarization
degree $p_\mathrm{n}$ of non-thermal intensity, which is a measure of the degree of field order.
$p_\mathrm{n}$ at \wave{6.2} remains roughly constant over the radial range 8--17\,kpc.
$p_\mathrm{n}$ at \wave{3.6} is similar, except for the radial range 8--9\,kpc,
which is affected by missing diffuse total intensity (see above).
$p_\mathrm{n}$ at \wave{3.6} is somewhat smaller than $p_\mathrm{n}$ at \wave{6.2} beyond 15\,kpc radius.
Such a behaviour is hard to explain because $p_\mathrm{n}$ should
decrease with increasing wavelength due to increasing Faraday depolarization.
Some large-scale polarized emission at \wave{3.6} seems to be missing at large radii.
Hence, polarization data at \wave{3.6} beyond 15\,kpc radius will not be included in the further
analysis.

$p_\mathrm{n}$ at \wave{11.3} and \wave{20.5} are almost constant with radius, with lower percentages
than at \wave{3.6} and \wave{6.2}, consistent with significant Faraday depolarization at those wavelengths.

\section{Diffusion of cosmic-ray electrons}
\label{sec:diff}

The map at \wave{20.5} shown in Figure~\ref{fig:cm3+20nth} (middle panel) looks like a
smoothed version of the map at \wave{3.6}, which is the result of CRE diffusion.
The diffusion length in spiral galaxies has been estimated by various authors.
By comparing the distributions of star formation surface density and radio continuum,
\citet{heesen19} found diffusion lengths at \wave{22} of 1.0--3.4\,kpc for three spiral galaxies.
With a similar method, \citet{vollmer20} derived diffusion lengths at \wave{21} of
0.80--2.4\,kpc for a sample of eight spiral galaxies. By comparing the distributions
of the thermal and non-thermal emission components of M~31, \citet{berkhuijsen13}
estimated a CRE diffusion length at \wave{20.5} of $l_\mathrm{diff,20.5}=(0.93\pm0.21)$\,kpc
in the sky plane. Applying the method of wavelet cross-correlation between radio and far-infrared
intensities, \citet{taba13b} derived $l_\mathrm{diff,20.5}=(0.73\pm0.09)$\,kpc in the sky plane.

The clumpiness of the emission torus at \wave{3.6} (Fig.~\ref{fig:cm3+20nth}, top panel) indicates that
CRE diffusion does not play a significant role at this wavelength and this resolution,
hence $l_\mathrm{diff,3.6} \lesssim 1\farcm5 \equiv 0.34$\,kpc in the sky plane.
To measure the diffusion length at \wave{20.5}, we smoothed the \wave{3.6} map
with Gaussian functions
to resolutions increasing in steps of $15\arcsec$ and correlated the intensities with the
intensities at \wave{20.5} in linear and log scales, both above three times the noise levels,
pixel by pixel.
The highest Pearson correlation coefficient of 0.695 (linear/linear) and 0.675 (log/log)
was found for the \wave{3.6} map smoothed to $2\farcm5$. The map at $2\farcm5$
(0.57\,kpc) shown in Figure~\ref{fig:cm3+20nth} (bottom panel) is most similar to the
\wave{20.5} map at $1\farcm5$ (Figure~\ref{fig:cm3+20nth}, middle panel). We conclude
that the average CRE diffusion length at \wave{20.5} is $l_\mathrm{diff,20.5} \simeq 0.57$\,kpc
in the sky plane,
similar to the estimate by \citet{taba13b}.

The diffusion length, $l_\mathrm{diff}$, depends on the diffusion coefficient, D, and the
CRE lifetime, $\tau$, as $l_\mathrm{diff}=(\mathrm{D} \cdot \tau)^{\,0.5}$.
\footnote{Instead, $l_\mathrm{diff}= 2 \, (\mathrm{D} \cdot \tau)^{0.5}$ was used by several authors
\citep{mulcahy14,vollmer20,nasirzadeh24}.}
If $\tau$ is limited by
losses due to synchrotron emission or Inverse Compton scattering
($\tau\propto E_\mathrm{CRE}^{\,-1} \propto \lambda^{\,0.5}$),
$l_\mathrm{diff}$ varies with wavelength $\lambda$ as $l_\mathrm{diff} \propto \lambda^{\,0.250}$
for constant D, or $l_\mathrm{diff} \propto \lambda^{\,0.375}$ for an energy-dependent
$\mathrm{D} \propto E_\mathrm{CRE}^{\,0.5}$ \citep{strong07}.
Our estimates of $l_\mathrm{diff,20.5}\simeq 0.57$\,kpc and $l_\mathrm{diff,3.6} \lesssim 0.34$\,kpc
in the sky plane yield an exponent of the wavelength dependence of $l_\mathrm{diff}$ of $\gtrsim0.3$,
which is consistent with energy-dependent D and either synchrotron or Inverse Compton losses.

The corresponding diffusion coefficient is $\mathrm{D}=l_\mathrm{diff}^{\,2} / \tau$, where $\tau$ is the
residence time of CREs emitting at \wave{20.5} in the torus. When assuming that $\tau$ is
dominated by synchrotron losses, $\tau \simeq 5 \cdot 10^7$ years in a magnetic field of
$\simeq 6.3\,\mu$G strength (see Section~\ref{sec:Btoteq}), we get $\mathrm{D}\simeq 2 \cdot 10^{27}$\,cm$^2$/s,
much smaller than the values of $\mathrm{D}\simeq$\,(3--6)\,$\cdot \,10^{28}$\,cm$^2$/s
favoured in models of CR propagation in the Galaxy \citep{moskalenko02}
and below the range of values derived for spiral galaxies
of $\mathrm{D}\simeq$\,(1.4--8.9)\,$\cdot \,10^{28}$\,cm$^2$/s
\citep{mulcahy16,heesen19,vollmer20}.
This result seems to be in conflict with the expectation that CRE diffusion is faster in a
galaxy with a strongly ordered field like that in M~31 \citep{taba13b,nasirzadeh24}.
With a residence time of CREs in the torus of a few times shorter than their lifetime
due to synchrotron losses, we would a achieve a diffusion coefficient typical to those
in other spiral galaxies.
Inverse Compton losses of CREs with the stellar radiation field in the emission
torus of M~31 may play an important role and significantly reduce the CRE lifetime.

\section{Magnetic field strengths in the emission torus}
\label{sec:results}

\subsection{Strength of the total equipartition field}
\label{sec:Btoteq}

The synchrotron intensity, $I_\mathrm{syn}$, scales with the density, $N_\mathrm{CRE}$, of CREs
in the energy range corresponding to the observation frequency and the total magnetic field strength,
$B_\mathrm{tot,\perp}$, in the sky plane as
$I_\mathrm{syn} \propto N_\mathrm{CRE} \, B_\mathrm{tot,\perp}^{\,\, (1 + \alphan)}$.
If no independent information about $N_\mathrm{CRE}$ is available, for example from $\gamma$ rays,
a relation between $N_\mathrm{CRE}$ and $B_\mathrm{tot}$ is needed to find $B_\mathrm{tot}$.
A widely used assumption is that of equipartition between the energy densities of the total
CRs and of the total magnetic field as developed by \citet{beck+krause05}, with
improvements by \citet{arbutina12}.

\begin{figure}[htbp]
\begin{center}
\includegraphics[width=0.95\columnwidth]{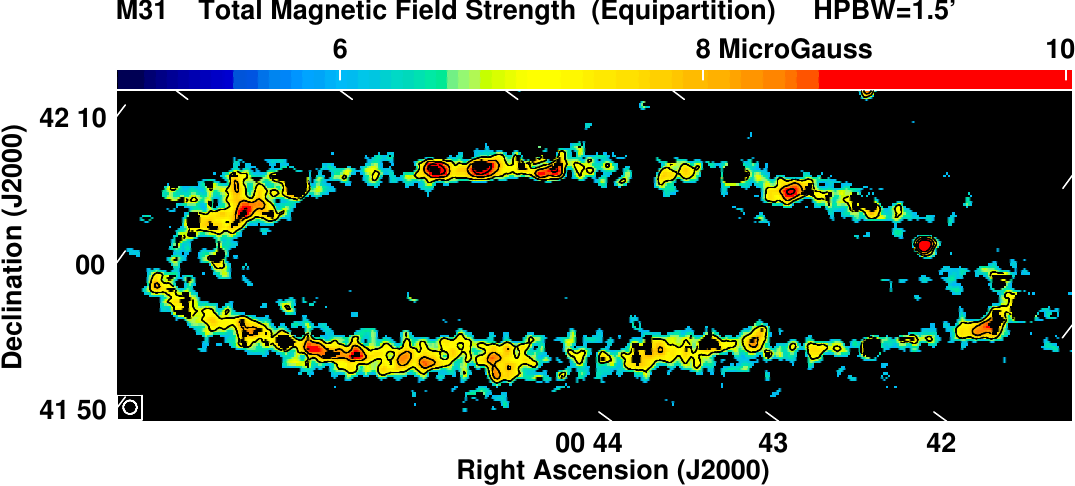}

\vspace{5mm}
\includegraphics[width=0.95\columnwidth]{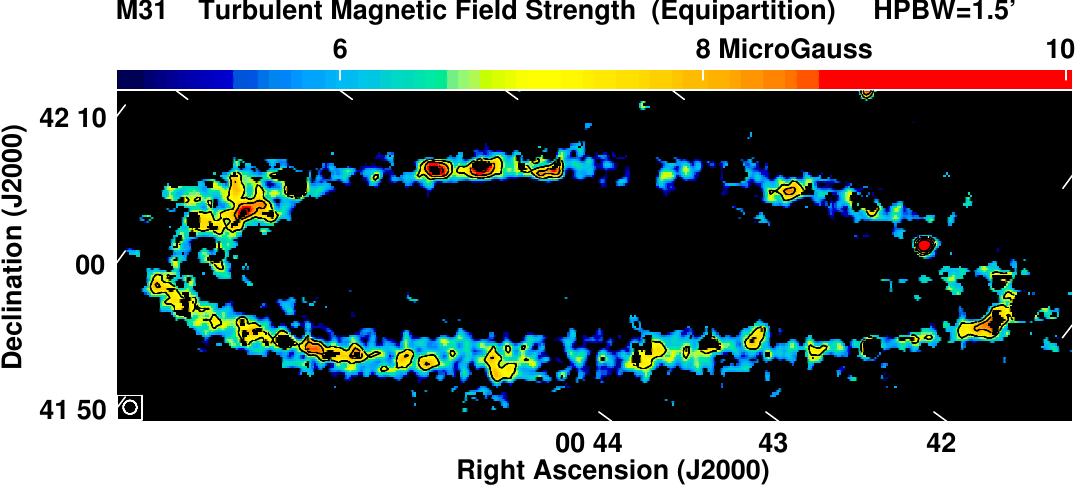}

\vspace{5mm}
\includegraphics[width=0.95\columnwidth]{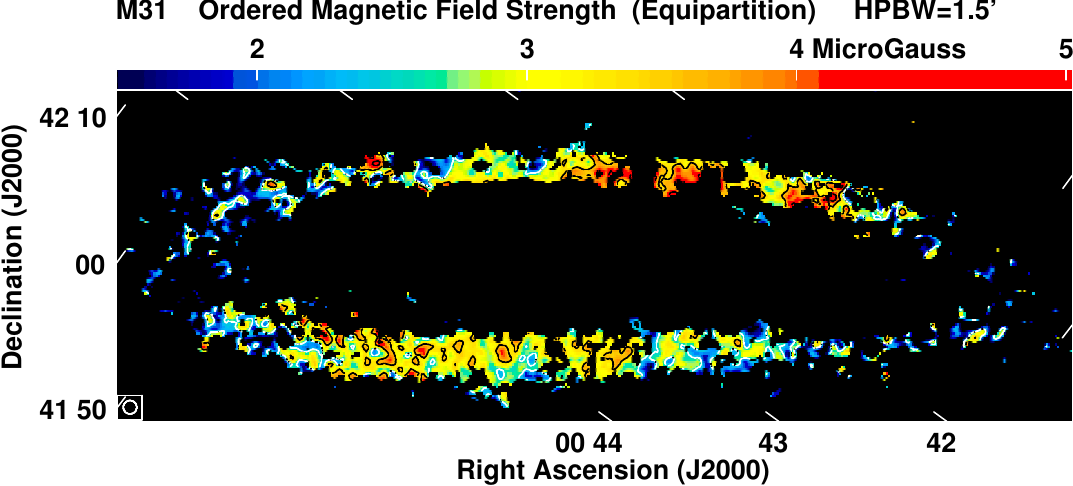}
\caption{Maps of the strengths (in $\mu$G) of the total (\textbf{top}), isotropic turbulent
(\textbf{middle}), and ordered magnetic fields in the sky plane (\textbf{bottom}) at $1\farcm5$ resolution,
computed from the total, unpolarized, and polarized non-thermal intensities at \wave{3.6}, assuming local
equipartition between the energy densities of total magnetic fields and total CRs.
Contour levels are at 7, 8, and 9\,$\mu$G (total field and turbulent field),
and 2.5, 3.5, and 4.5\,$\mu$G (ordered field).
Outside of the emission torus, synchrotron spectral indices are uncertain, so that no field
strengths were computed.
}
\label{fig:Bmaps}
\end{center}
\end{figure}

Equilibrium between magnetic and cosmic-ray energy densities is reasonable at least on the scale
of a large spiral galaxy because both quantities are closely related to the rate of star
formation. Violation of overall equipartition may be caused by bursts of star formation or massive
interaction with a neighbouring galaxy, both of which do not apply to M~31.

\citet{seta19} argued that the equipartition assumption is not valid below scales of the CRE propagation
length. In Section~\ref{sec:diff}, we estimated an isotropic diffusion length at \wave{3.6} of
$l_\mathrm{diff,3.6} \lesssim 0.34$\,kpc, so that applying the equipartition
assumption to our synchrotron map at this wavelength is reasonable.

The equipartition assumption yields $N_\mathrm{CRE} \propto N_\mathrm{CR} \propto B_\mathrm{tot}^2$.
In the ISM, protons dominate the total CRs. Hence,
the CR energy density is determined by integrating over the energy spectrum of the protons.
This allows us to calculate the total magnetic field strength from the synchrotron intensity,
$I_\mathrm{syn}$, according to \citet{beck+krause05}:
\begin{equation}
B_\mathrm{tot,\perp} \propto [\, I_\mathrm{syn} \, (K + 1) \, / \, L_\mathrm{syn} \,]^{\, 1/(3 + \alphan)}\, ,
\label{eq:equi}
\end{equation}

\noindent where $\alphan$ is the synchrotron spectral index and $L_\mathrm{syn}$ is the effective pathlength
through the synchrotron-emitting source. $K$ is the ratio of number densities of CR protons and electrons in the
relevant energy range. For diffusive shock acceleration of CRs in the ISM and moderate
shock strengths, $K\simeq100$ is a reasonable assumption in the energy range where the spectra
of protons and electrons have a similar spectral index, i.e. beyond the rest-mass energy of protons
\citep{bell78} and below the energy where losses of CR electrons become important.

The degree of non-thermal polarization, $p_\mathrm{n}$, is needed for the geometrical correction to compute
$B_\mathrm{tot}$ from $B_\mathrm{tot,\perp}$ \citep{beck+krause05}. As the equipartition strength is only
a weak function of $p_\mathrm{n}$, we used the average value of $p_\mathrm{n}=18\%$ at \wave{3.6} for the
radial ring 8--12\,kpc (see Figure~\ref{fig:radial}, bottom panel) for all pixels in the map.

The emission torus of M~31 can be described by an elliptical profile with a full width, $W$, in the galaxy
plane along the line of sight at $90\degr$ inclination and a full height, $H$, perpendicular to the
galaxy plane. The pathlength, $L$, through the torus seen under the inclination angle, $i$, is
\begin{equation}
L \, = H \, / \, [\, 1 - \sin^2 i \, ( 1 - (H/W)^2)\, ]^{\, 0.5} \, .
\label{eq:path}
\end{equation}

The radial profile of synchrotron emission in Figure~\ref{fig:radial_I} yields a torus width of
$W_\mathrm{syn}\simeq4.5$\,kpc in the galaxy plane. This is also the pathlength through the torus along the line of
sight on the minor axis at $90\degr$ inclination, which increases to about 15\,kpc on the major axis.
\citet{berkhuijsen13} found an exponential scale height of the non-thermal emission at \wave{20.5}
of $H_\mathrm{syn} \simeq0.33$\,kpc, which is significantly smaller than the average scale height
of 1.4\,kpc at \wave{20} for a sample of edge-on galaxies \citep{krause18}.
The full height of $2H_\mathrm{syn} \simeq 0.66$\,kpc means that the synchrotron-emitting torus of M~31
is flat ($H_\mathrm{syn}/W_\mathrm{syn} \ll 1$) and Eq.~(\ref{eq:path}) simplifies to
$L_\mathrm{syn} = H_\mathrm{syn} \, / \cos i \simeq 2.5$\,kpc.

A pixel by pixel map of $B_\mathrm{tot}$ in the emission torus was computed from the maps at $1\farcm5$
resolution of the total synchrotron intensities $I_\mathrm{syn}$ at \wave{3.6} (Fig.~\ref{fig:cm3+20nth}, top panel)
and the spectral index map $\alphan$ between \wave{3.6} and \wave{20.5} (corrected for AME) that
is reliable in the emission torus (8--12\,kpc radial range). Only pixels were used where
$0.53 \le \alphan \le 1.4$, which is the range where the equipartition assumption is applicable.
For $\alphan < 0.53$, corresponding to spectral indices $\epsilon$ of the CRE energy spectrum of
$\epsilon < 2.06$, the total CR energy density is dominated by the largest CRE energies, while for $\alphan > 1.4$
or $\epsilon > 3.8$, the CRE energy spectrum is dominated by synchrotron and/or Inverse Compton losses,
so that $K\simeq100$ is no longer valid \citep[see Appendix in ][]{heesen23}.
We also discarded the central region of M~31 within 6\,kpc radius for two reasons:
the spectral index is uncertain in this region (see Fig.~\ref{fig:radial}, top panel)
and the pathlength is different from that through the emission torus.

Figure~\ref{fig:Bmaps} (top panel) shows the strength of the total equipartition
field $B_\mathrm{tot}$. The mean total field strength in the torus, averaged over sectors of $10\degr$
azimuthal width in the radial ring 8--12\,kpc, is $(6.8\pm0.4)\,\mu$G. Due to the nonlinear relation
between total field strength and synchrotron intensity, this value is biased towards high peaks.
The total equipartition field strength based on the mean synchrotron intensity and the
mean synchrotron spectral index in the torus is more reliable and yields $(6.3\pm0.2)\,\mu$G.

The total field is decomposed into the isotropic turbulent field and the ordered field,
$B_\mathrm{tot}^2 = B_\mathrm{turb}^2 + B_\mathrm{ord}^2$.
The ordered field has two components, the regular (mean) field and the anisotropic turbulent field,
$B_\mathrm{ord}^2 = B_\mathrm{reg}^2 + B_\mathrm{an}^2$.
The strength of the isotropic turbulent field is estimated in Section~\ref{sec:Bturbeq}, that
of the ordered field in Section~\ref{sec:Bordeq}, that of the regular field in Section~\ref{sec:Breg},
and that of the anisotropic turbulent field in Section~\ref{sec:Ban}.

\subsection{Strength of the isotropic turbulent equipartition field}
\label{sec:Bturbeq}

Applying the equipartition assumption to the unpolarized intensity, $I_\mathrm{unpol}$,
emitted by CREs in an isotropic turbulent field $B_\mathrm{turb,\perp}$ in the sky plane yields
\begin{eqnarray}
I_\mathrm{unpol} & \propto & N_\mathrm{CRE} \, B_\mathrm{turb,\perp}^{\, (1 + \alphan)} \nonumber \\
& \propto & B_\mathrm{tot}^2 \, B_\mathrm{turb,\perp}^{\, (1 + \alphan)}  \nonumber \\
& \propto & B_\mathrm{turb}^2 \, B_\mathrm{turb,\perp}^{\, (1 + \alphan)} \, + \, B_\mathrm{ord}^2 \, B_\mathrm{turb,\perp}^{\, (1 + \alphan)}\, .
\label{eq:unpol1}
\end{eqnarray}

\noindent As the degrees of polarization of non-thermal emission are relatively small, we may assume
$B_\mathrm{ord}^2 \ll B_\mathrm{turb}^2$, so that Equation~(\ref{eq:unpol1}) simplifies to
\begin{equation}
I_\mathrm{unpol}  \propto B_\mathrm{turb}^2 \, B_\mathrm{turb,\perp}^{\, (1 + \alphan)}\, ,
\label{eq:unpol2}
\end{equation}

\noindent with $B_\mathrm{turb,\perp}^2 = 2/3 \, B_\mathrm{turb}^2$.
Hence, the equipartition estimate can also be applied to derive the strength of $B_\mathrm{turb}$ from
the unpolarized synchrotron intensity, $I_\mathrm{unpol}$, which follows from
\begin{equation}
I_\mathrm{unpol}  = I_\mathrm{syn} \, - \, I_\mathrm{pol}/p_\mathrm{0} \, ,
\label{eq:unpol3}
\end{equation}

\noindent where $I_\mathrm{pol}$ is the polarized intensity and $p_\mathrm{0}$ is the intrinsic
degree of polarization, $p_\mathrm{0} = 73.5\%$, for the average non-thermal spectral index of $\alphan=0.85$.

Figure~\ref{fig:Bmaps} (middle panel) shows the resulting map of $B_\mathrm{turb}$.
The equipartition strength based on the mean unpolarized synchrotron intensity
in the torus is $(5.4\pm0.2)\,\mu$G.

\subsection{Strength of the ordered equipartition field}
\label{sec:Bordeq}

Finally, the linearly polarized emission, $I_\mathrm{pol}$, is related to the ordered equipartition field strength
$B_\mathrm{ord,\perp}$ in the sky plane:
\begin{eqnarray}
I_\mathrm{pol} & \propto & N_\mathrm{CRE} \, B_\mathrm{ord,\perp}^{\, (1 + \alphan)} \nonumber \\
& \propto & B_\mathrm{tot}^2 \, B_\mathrm{ord,\perp}^{\, (1 + \alphan)}  \nonumber \\
& \propto & B_\mathrm{turb}^2 \, B_\mathrm{ord,\perp}^{\, (1 + \alphan)} \, + \, B_\mathrm{ord}^2 \, B_\mathrm{turb,\perp}^{\, (1 + \alphan)} \, .
\label{eq:pol1}
\end{eqnarray}

\noindent With $B_\mathrm{ord}^2 \ll B_\mathrm{turb}^2$, Equation~\ref{eq:pol1} simplifies to
\begin{equation}
I_\mathrm{pol}  \propto B_\mathrm{turb}^2 \, B_\mathrm{ord,\perp}^{\, (1 + \alphan)}\, .
\label{eq:pol2}
\end{equation}

\noindent Equation~(\ref{eq:pol2}) demonstrates that $B_\mathrm{ord,\perp}$
cannot be estimated from $I_\mathrm{pol}$ using the equipartition formula.
Instead, $B_\mathrm{ord,\perp}$ can be computed via the relation
$B_\mathrm{ord,\perp}^2 = B_\mathrm{tot}^2 - B_\mathrm{turb,\perp}^2$.
The resulting map is shown in Fig.~\ref{fig:Bmaps} (bottom panel).
The equipartition strength based on the mean polarized synchrotron
intensity in the torus is $(3.2\pm0.3)\,\mu$G.

\subsection{Uncertainties in the equipartition field strengths}
\label{sec:errors}

The equipartition field strengths $B$ shown in Figure~\ref{fig:Bmaps} are subject to various uncertainties:

(1) To be conservative, we assume that the parameters $K$ and $L_\mathrm{syn}$ are known only to
$\lesssim50\%$ accuracy, which leads to a total systematical uncertainty in $B$ of $\lesssim20\%$.

(2) The root mean square (rms) noise $\sigma$ of the radio intensity maps causes statistical uncertainties. At the
lowest level of $5\sigma$ the maximum uncertainty in $B$ is $\simeq20\%$.

(3) Another statistical uncertainty emerges from that in $\alphan$. An uncertainty of
$\delta\alphan=0.1$ leads to an uncertainty in $B$ of $\simeq7\%$ for $\alphan=0.7$,
$\simeq8\%$ for $\alphan=0.8$, and $\simeq10\%$ for $\alphan\ge1.0$.

(4) The volume filling factor $f_\mathrm{syn}$ of synchrotron emission may be smaller than one,
so that the pathlength $L_\mathrm{syn}$ has to be replaced by $L_\mathrm{syn} \cdot f_\mathrm{syn}$,
which increases the average total field strength $B_\mathrm{tot,\perp}$.

(5) Fluctuations of $B$ within the telescope beam and/or along the line of sight lead
to an overestimate of the equipartition value. Due to the highly nonlinear dependence of $I_\mathrm{syn}$ on $B_\mathrm{tot,\perp}$ (Sect.~\ref{sec:Btoteq}),
the average equipartition value $\langle B_\mathrm{tot,\perp} \rangle$ derived from synchrotron intensity is biased
towards large field strengths and hence is an overestimate if $B_\mathrm{tot}$ varies along the line
of sight and/or across the telescope beam \citep{beck03}.

For $\alpha_\mathrm{n}$\,=\,1 and constant density of CRE (the non-equipartition case), the overestimation factor
$g$ of the total field is
\begin{equation}
g = \langle B_\mathrm{tot,\perp}^{\,\,2} \rangle^{1/2} \, / \langle B_\mathrm{tot,\perp} \rangle \,\, = (1 + q^2)^{1/2}\, ,
\label{eq:over1}
\end{equation}

\noindent while for $\alpha_\mathrm{n}$\,=\,1 and the equipartition case the overestimation factor, $g$, of the total field
\citep[Appendix A in][]{stepanov14} is
\begin{equation}
g = \langle B_\mathrm{tot,\perp}^{\,\,4} \rangle^{1/4} \, / \langle B_\mathrm{tot,\perp} \rangle \,\, =
(1 + {8\over3} q^2 + {8\over9} q^4)^{1/4}\, ,
\label{eq:over2}
\end{equation}

\noindent where $\langle \,\, \rangle$ indicates the volume average along the line of sight and across the beam, and
$q=\langle \delta B_\mathrm{tot,\perp}^2 \rangle^{1/2}\, / \langle B_\mathrm{tot,\perp} \rangle$ is the
amplitude of the field fluctuations relative to the average field.
For extreme fluctuations of $q=1$, the overestimation factor is 1.41 and 1.46 for the
non-equipartition and equipartition cases, respectively.
As equipartition is not valid on the spatial scale of field fluctuations,
Equation~(\ref{eq:over1}) is the more realistic case.

The dispersion in $B_\mathrm{tot}$ in Figure~\ref{fig:Bmaps} (top panel), determined in
several regions of a few beam sizes extent, serves as an estimate of the 2D
field fluctuations in the sky plane on the scale of our spatial resolution of 0.34\,pc and is measured to be
$q_\mathrm{2D} = 0.13\pm0.02$. Assuming isotropic turbulence yields
$q_\mathrm{3D} = 3/2 \, q_\mathrm{2D} = 0.20\pm0.03$ in 3D. As our spatial resolution
is close to the largest scale of the Kolmogorov power spectrum of small-scale fields, the dispersion
is representative for the total small-scale field (Amit Seta, private communication).
$q_\mathrm{3D} \simeq0.2$ gives overestimation factors of $g \simeq 1.02$ and $g \simeq 1.03$
for the non-equipartition and equipartition cases, respectively.

The ratio $q_\mathrm{3D}$ is related to the volume filling factor $f_\mathrm{B}$ of the total field via
\begin{eqnarray}
\langle \delta B_\mathrm{tot,\perp}^2 \rangle & = & \langle B_\mathrm{tot,\perp} \rangle^2 + \langle \delta B_\mathrm{tot,\perp}^2 \rangle \nonumber \\
& = & \langle B_\mathrm{tot,\perp} \rangle^2 /\, f_\mathrm{B} \nonumber\, ,
\end{eqnarray}
\begin{equation}
f_\mathrm{B} = 1 \, / \,(q_\mathrm{3D}^2+1)\, .
\end{equation}

With $q_\mathrm{3D} \simeq 0.2$, we get $f_\mathrm{B}\simeq 0.96$.
To our knowledge, this is the first estimate of the volume filling factor of the magnetic field
in the diffuse ISM
based on observations.\\

The Bayesian approach presented by \citet{zychowicz25}
allows one to get a better handle on the uncertainties of the input parameters.
A further development would be to allow curved radio spectra.

\subsection{Strength of the non-equipartition total field}
\label{sec:Btotneq}

The maps shown in Figure~\ref{fig:Bmaps} are based on the assumption of equipartition at the scale of
our spatial resolution of $\simeq 0.34\,\mathrm{kpc} \times 1.31\,\mathrm{kpc}$ in the galaxy plane
along the major and the minor axis, respectively. The average diffusion length of $l_\mathrm{diff,20.5}
\simeq 0.57$\,kpc, estimated in Section~\ref{sec:diff}, indicates that the equipartition assumption
is not valid at \wave{20.5} on the scale of our spatial resolution.

\begin{figure}[htbp]
\begin{center}
\includegraphics[width=0.95\columnwidth]{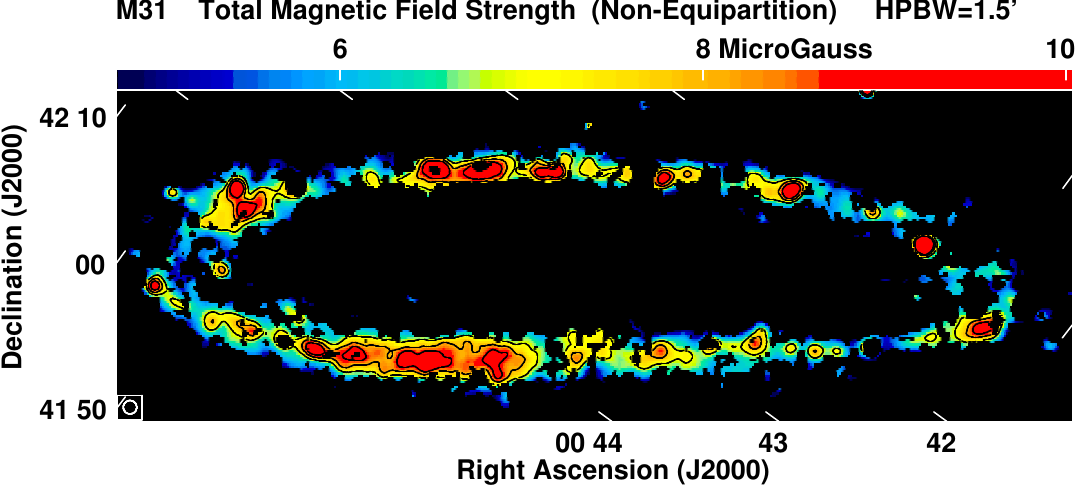}
\hfill
\caption{Map of the strength of the total magnetic field (in $\mu$G), determined from the total
non-thermal intensities at \wave{20.5} at $1\farcm5$ resolution, assuming global (but not local) equipartition
between the energy densities of total magnetic fields and total CRs. Contour levels are
at 7, 8, and 9\,$\mu$G. The colour scale is the same as that in Fig.~\ref{fig:Bmaps} (top panel).}
\label{fig:Btot_noneq}
\end{center}
\end{figure}

\begin{figure}[htbp]
\begin{center}
\includegraphics[width=0.95\columnwidth]{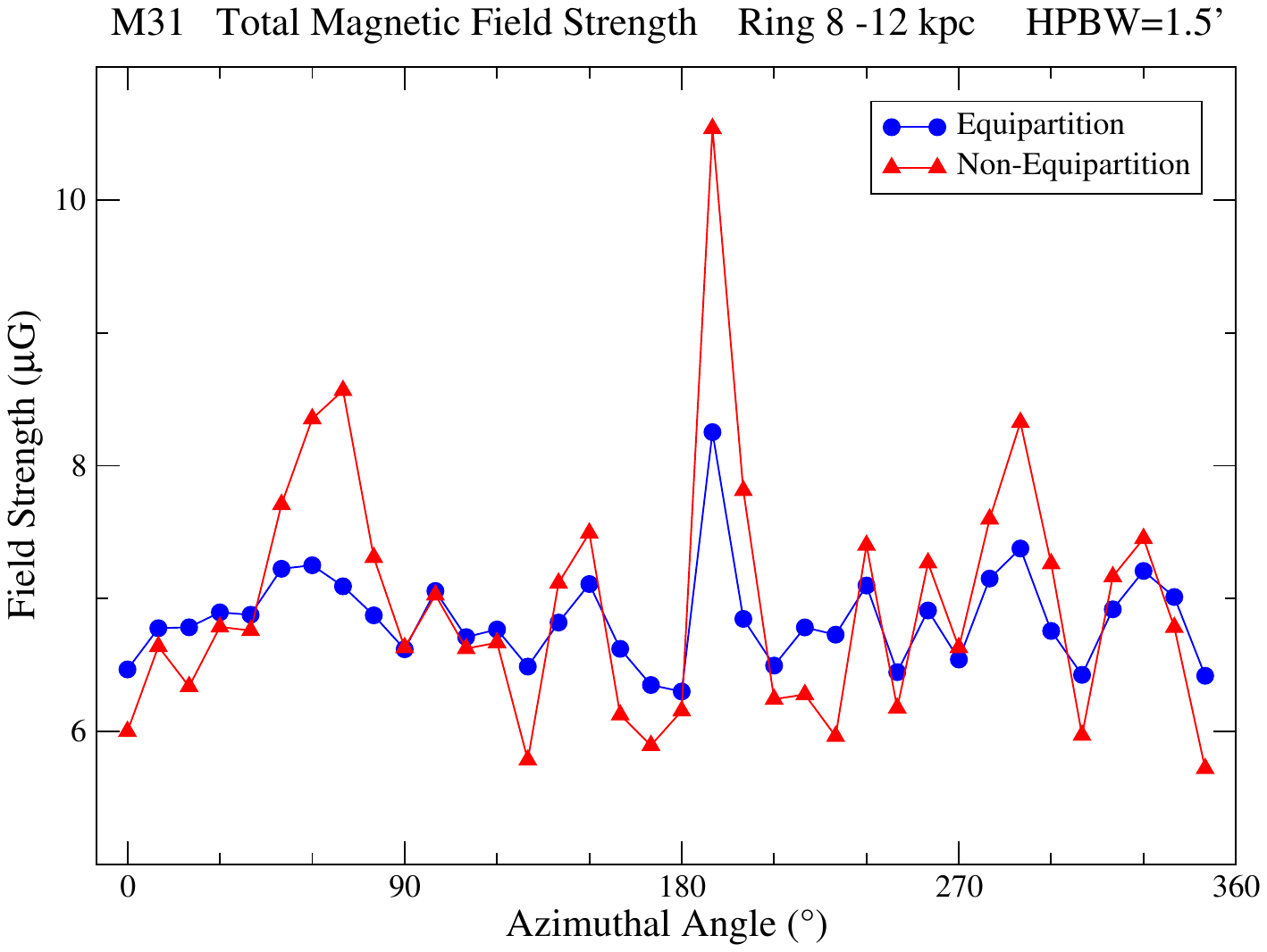}
\hfill
\caption{Variations in the strengths of the equipartition and non-equipartition total fields
 at $1\farcm5$ resolution with azimuthal angle in the galaxy plane, counted from the north-eastern
 major axis in the sky plane (left side in Fig.~\ref{fig:Btot_noneq}), averaged in $10\degr$--sectors
 of the radial ring 8--12\,kpc.}
\label{fig:Btot_azimuthal}
\end{center}
\end{figure}

The CR diffusion is anisotropic \citep{chuvilgin93,yan04} and expected to be faster in a highly ordered
magnetic field \citep{taba13b,nasirzadeh24}, like that in the emission torus of M~31.
The diffusion length along the torus may well be larger than the average isotropic one of
$\lesssim0.48$\,kpc obtained in Section~\ref{sec:diff}. Furthermore,
\citet{tharakkal23b} concluded from their MHD model of the Parker instability that
there are no signs of equipartition at kiloparsec scales. Signatures of Parker instabilities
in M~31 were indeed observed \citep{beck89}.

In the following, we consider the extreme case of rapid CRE diffusion, ignoring energy losses and escape,
resulting in constant CRE density along the torus, hence on a scale of several 10\,kpc.
Then, variations in $I_\mathrm{syn}$ are solely caused by variations in the total magnetic field strength
$B_\mathrm{tot,\perp}$:
\begin{equation}
B_\mathrm{tot,\perp} \propto [\, I_\mathrm{syn} \, (K + 1) \, / \, L \,]^{\, 1/(1 + \alphan)} \, .
\label{eq:nonequi}
\end{equation}

The non-equipartition map of $B_\mathrm{tot}$ (Figure~\ref{fig:Btot_noneq}) was computed from the maps of
total intensities $I_\mathrm{syn}$ at \wave{20.5} (Fig.~\ref{fig:cm3+20nth}, middle panel) and the
spectral index map $\alphan$ from \citet{beck20}, corrected for AME. To calibrate the overall level
of $B_\mathrm{tot,\perp}$, a minimal assumption was required, i.e. that equipartition is valid on
average in the emission torus.

The non-equipartition total field presented in Figure~\ref{fig:Btot_noneq} shows larger
variations compared to the equipartition total field (Fig.~\ref{fig:Bmaps}, top panel), as expected.
The largest field strengths occur in star-forming regions with a maximum of 16.4\,$\mu$G,
compared to 10.6\,$\mu$G for the equipartition field.
This is illustrated in Figure~\ref{fig:Btot_azimuthal} where equipartition and
non-equipartition total field strengths are compared along azimuthal angle in the radial ring 8--12\,kpc.
The mean non-equipartition field strength in the torus, averaged in sectors of $10\degr$ azimuthal width
in the ring 8--12\,kpc, is $(7.0\pm1.0)\,\mu$G, hence similar but with a larger dispersion compared to
the equipartition field.

\subsection{Strength of the regular field}
\label{sec:Breg}

The ordered field $B_\mathrm{ord,\perp}$ in the sky plane presented in Fig.~\ref{fig:Bmaps} (bottom panel)
has two components, the regular field $B_\mathrm{reg,\perp}$ and the anisotropic turbulent field $B_\mathrm{an,\perp}$.
A separation of these two components is not possible based on polarized emission alone but needs additional
data from RMs that are a tracer of regular fields along the line of sight.
Faraday rotation is an ideal tool to study the structure of regular fields because it is insensitive to
anisotropic turbulent fields. The mean-field dynamo generates large-scale modes of the regular field
with spiral structures that are visible as large-scale RM patterns. Gas motions may tangle the regular field
and distort the RM pattern.
As discussed in \citet{beck20}, $B_\mathrm{reg}$ and $B_\mathrm{an}$ may have spiral structures with different
pitch angles that are shaped by different physical processes.

\begin{figure}[htbp]
\begin{center}
\includegraphics[width=0.95\columnwidth]{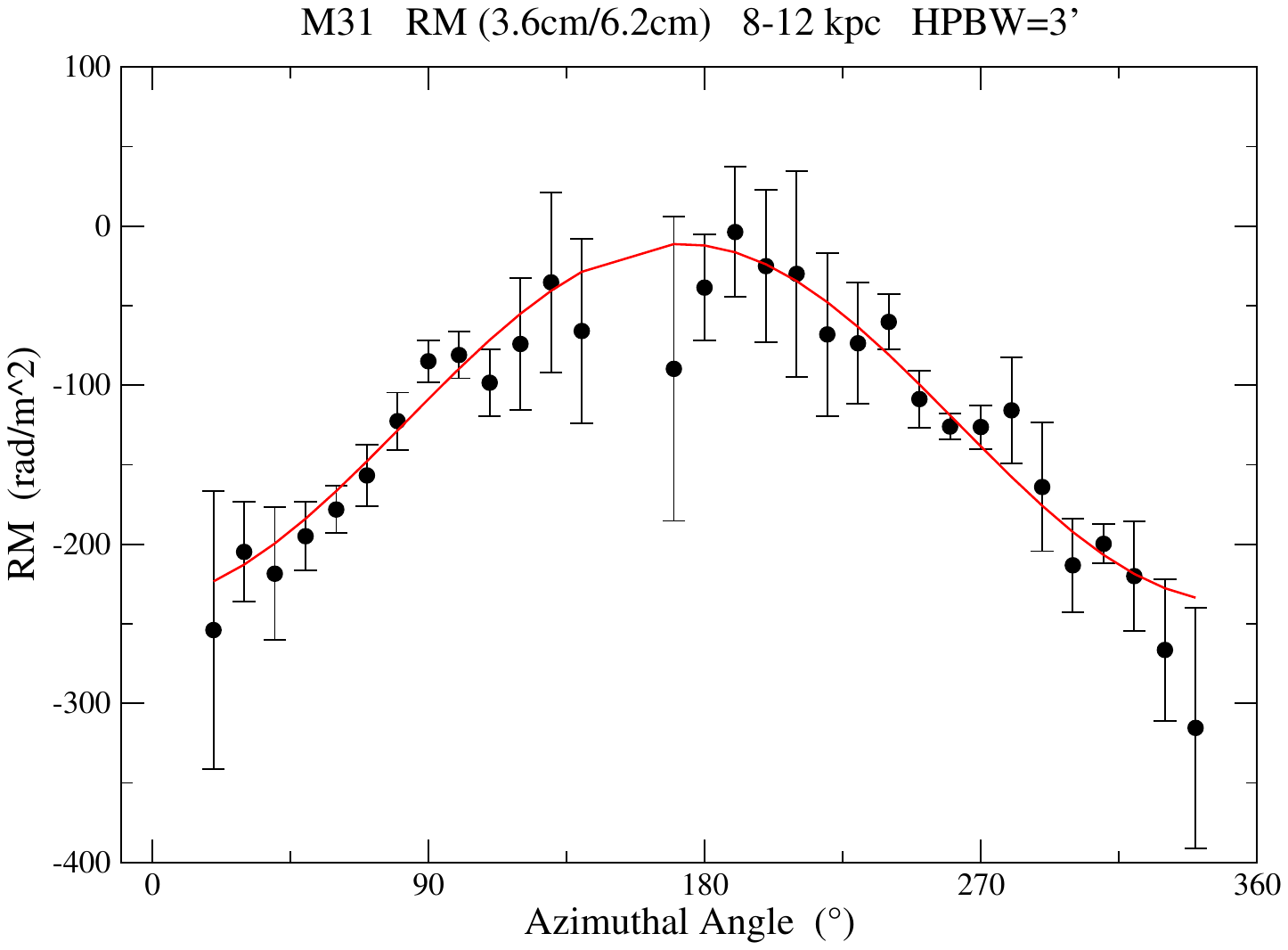}
\hfill
\caption{Variation in RMs between \wave{3.6} and
\wave{6.2} at $3\arcmin$ resolution with azimuthal angle in the galaxy plane,
averaged in $10\degr$--sectors of the radial ring 8--12\,kpc, and fitted by a
sinusoidal line (in red).
}
\label{fig:rm3_6_azm}
\end{center}
\end{figure}

In a perfectly ASS regular field, the basic ($m=0$) dynamo mode,
RM is expected to vary with azimuthal angle \citep{krause89a} as
\begin{equation}
\mathrm{RM} = \mathrm{RM}_\mathrm{fg} + \mathrm{RM}_0 \, \cos\,(\phi - \xi_\mathrm{reg,\parallel})\, ,
\label{eq:RM}
\end{equation}

\noindent where $\mathrm{RM}_\mathrm{fg}$ is the RM contribution from the Milky Way in the
foreground of M~31, $\phi$ is the azimuthal angle in the galaxy plane, and the phase
shift, $\xi_\mathrm{reg,\parallel}$, is the spiral pitch angle of the ASS field,
expected to be constant along $\phi$. $\mathrm{RM}_0$ is the RM amplitude of
the ASS mode near the north-eastern major axis of the projected emission torus,
i.e. at the azimuthal angle $\phi=0\degr+\xi_\mathrm{reg,\parallel}$.

The variation in RM with azimuthal angle in the galaxy plane is close to sinusoidal
\citep{beck82,berkhuijsen03,beck20}. Figure~\ref{fig:rm3_6_azm} shows the average
RMs at $3\arcmin$ resolution in the radial ring 8--12\,kpc. The cosine function
gives an excellent fit to the data,
yielding $\mathrm{RM}_\mathrm{fg}=(-124\pm5)\,\radm$ and amplitude $\mathrm{RM}_0=(-113\pm10)\,\radm$.
The negative sign of $\mathrm{RM}_0$ means that the regular field points away from us near the north-eastern
major axis (azimuthal angle $0\degr$).

The pitch angle of $\xi_\mathrm{reg,\parallel}=-8\degr\pm3\degr$ obtained from the fit
is larger than that of $-15\degr\pm4\degr$ measured by \citet{fletcher04} for the radial range
9--13\,kpc (at 780\,kpc distance). The probable reason is that the method applied by
\citet{fletcher04} is based on the position angles of the ordered fields, which are affected by
anisotropic turbulent fields that have a smaller pitch angle \citep{beck20} (see also Section~\ref{sec:Ban}).

The observed amplitude, $\mathrm{RM}_0$, is given by
\begin{equation}
\mathrm{RM}_0 = 0.5 \cdot 0.812 \cdot B_\mathrm{reg} \, \sin i \, \langle n_\mathrm{e} \rangle \, L_\mathrm{th}\, ,
\label{eq:Breg1}
\end{equation}

\noindent where $B_\mathrm{reg}$ (in $\mu$G) is the volume-averaged strength of the regular field,
$i$ is the galaxy's inclination,
$B_\mathrm{reg} \sin i$ is the component $B_\mathrm{reg,\parallel}$ of the average regular field along
the line of sight, $\langle n_\mathrm{e} \rangle$ is the average thermal electron density (in cm$^{-3}$),
and $L_\mathrm{th}$ (in parsecs) is the pathlength through the ionized medium. The factor 0.5
accounts for the fact that the observed average RM is smaller than the total RM from the far
side of the emission region by this factor, assuming symmetric distributions of $B_\mathrm{reg}$ and
$\langle n_\mathrm{e} \rangle$ along the line of sight.

Equation~(\ref{eq:Breg1}) allows one to compute the strength of the regular field $B_\mathrm{reg}$ if
$L_\mathrm{th}$ and $\langle n_\mathrm{e} \rangle$ are known. An exponential scale height of the warm thermal
gas of $H_\mathrm{th,exp}=(0.55\pm0.1)$\,kpc in the radial range 9--11\,kpc (at 780\,kpc distance)
was obtained from fitting the Faraday depolarization data at \wave{6.2} and \wave{20.5} with
$3\arcmin$ resolution \citep{fletcher04}. We adopt a full thickness of $H_\mathrm{th}=(1.1\pm0.2)$\,kpc.
The full width of the torus of thermal gas in the galaxy plane of $W_\mathrm{th}\simeq3$\,kpc
(Fig.~\ref{fig:radial_I}) is smaller than that of the synchrotron emission, giving a pathlength through
the torus at $90\degr$ inclination of 3\,kpc on the minor axis and about 10\,kpc on the major axis.
The torus of thermal emission has a thicker profile than that of non-thermal emission.
As the amplitude $\mathrm{RM}_0$ is dominated by the pathlength around the major axis, $H \ll W$ is still valid,
and, according to Eq.~(\ref{eq:path}), the average pathlength at $75\degr$ inclination becomes
$L_\mathrm{th}\simeq(4.25\pm0.77)$\,kpc.

$\langle n_\mathrm{e} \rangle$ is a quantity that is hard to measure. The typical emission measure (EM) of H$\alpha$ emission
from diffuse warm ionized gas in the torus of M~31 of 15\,pc\,cm$^{-6}$ \citep{walterbos94}, a filling factor
of 0.2, and the above pathlength of 4.25\,kpc yields $\langle n_\mathrm{e} \rangle\,\simeq0.026$\,cm$^{-3}$.
\citet{beck19} estimated $\langle n_\mathrm{e} \rangle\,\simeq 0.033$\,cm$^{-3}$ from the total field strength
$B_\mathrm{tot}$ based on the relation $B_\mathrm{tot} \propto \mathrm{SFR}^{\,0.34}$ \citep{taba17}.

We applied a more direct method with the help of the thermal radio emission (Fig.~\ref{fig:radial_I}),
smoothed to $3\arcmin$, the same resolution as the RM data.
The thermal intensity, $I_\mathrm{th}$ (in mJy per beam), was transformed into the emission
measure EM (in cm$^{-6}$ pc) via \citep{ehle93,dickinson03}:
\begin{equation}
\mathrm{EM} = 4.12 \, I_\mathrm{th} \, T_\mathrm{e}^{0.35} \, \nu^{0.1} \, \theta^{-2} \, a^{-1} / \, 1.08\, ,
\label{eq:EM}
\end{equation}

\noindent where $T_\mathrm{e}$ is the electron temperature (in Kelvin), $\nu$ is the observation frequency (in GHz),
and $\theta$ is the telescope beamsize (in arcminutes). $a$ is a correction factor depending on electron temperature
and frequency \citep[see Table~3 in][]{dickinson03}, and the factor 1.08 accounts for the contribution of helium atoms
\citep{dickinson03}.
We assume that $T_\mathrm{e}=(7000\pm1000)$\,K \citep[see Fig.~1 in ][]{taba13b}, $\theta=3\arcmin$, and $a=0.98$.
$I_\mathrm{th}=(2.02\pm0.06)$\,mJy/beam at $\nu=8.35$\,GHz in the radial range 8--12\,kpc
yields $\mathrm{EM} = (24.0\pm1.2)$\,cm$^{-6}$\,pc, which is about twice larger than the average EM derived from
H$\alpha$ emission \citep{walterbos94} that is affected by absorption. Finally, $\langle n_\mathrm{e} \rangle$ follows from
\begin{equation}
\langle n_\mathrm{e} \rangle = (\mathrm{EM} \,f_\mathrm{th} / \, L_\mathrm{th})^{0.5}\, ,
\label{eq:ne}
\end{equation}

\noindent where $L_\mathrm{th}$ is the pathlength and $f_\mathrm{th}$ is the volume filling factor
of the diffuse warm ionized gas. $f_\mathrm{th}$ was measured so far only in the Milky Way with
help of pulsar dispersion measures and H$\alpha$ emission measures. \citet{berkhuijsen06} found
$f_\mathrm{th}$ = 0.05--0.3 at heights of 0--1\,kpc from the mid-plane. Due to the lack of data on M~31,
we adopt $f_\mathrm{th}=0.20\pm0.05$.
For the radial range 8--12\,kpc, we obtain $\langle n_\mathrm{e} \rangle \,= (0.034\pm0.005)$\,cm$^{-3}$, similar
to the previous estimates.

The strength of the regular field follows from
\begin{equation}
B_\mathrm{reg} = \mathrm{RM}_0\, / \,\,[\,0.5 \cdot 0.812 \cdot \sin i \, (L_\mathrm{th} \, \mathrm{EM} \, f_\mathrm{th})^{0.5} \,] \, .
\label{eq:Breg3}
\end{equation}

\noindent With $\mathrm{RM}_0=(-113\pm10)\,\radm$, Eq.~(\ref{eq:Breg3}) yields $B_\mathrm{reg}=(2.0\pm0.5)\,\mu$G.
The uncertainty in $B_\mathrm{reg}$ includes the noise error in EM (from the noise error in $I_\mathrm{th}$)
and the systematic uncertainties in $f_\mathrm{th}$ and $L_\mathrm{th}$.
The radial variation in $B_\mathrm{reg}$ is discussed in Section~\ref{sec:radial}.

\subsection{Strength of the anisotropic turbulent field}
\label{sec:Ban}

To compute the anisotropic turbulent field, we subtracted the regular field in the sky plane
from the observed ordered field in the sky plane. From Sect.~\ref{sec:Breg} we know that the
regular field is homogeneous along azimuthal angle in the radial range 8--12\,kpc and
has the strength $B_\mathrm{reg}$.

\begin{figure}[htbp]
\begin{center}
\includegraphics[width=0.95\columnwidth]{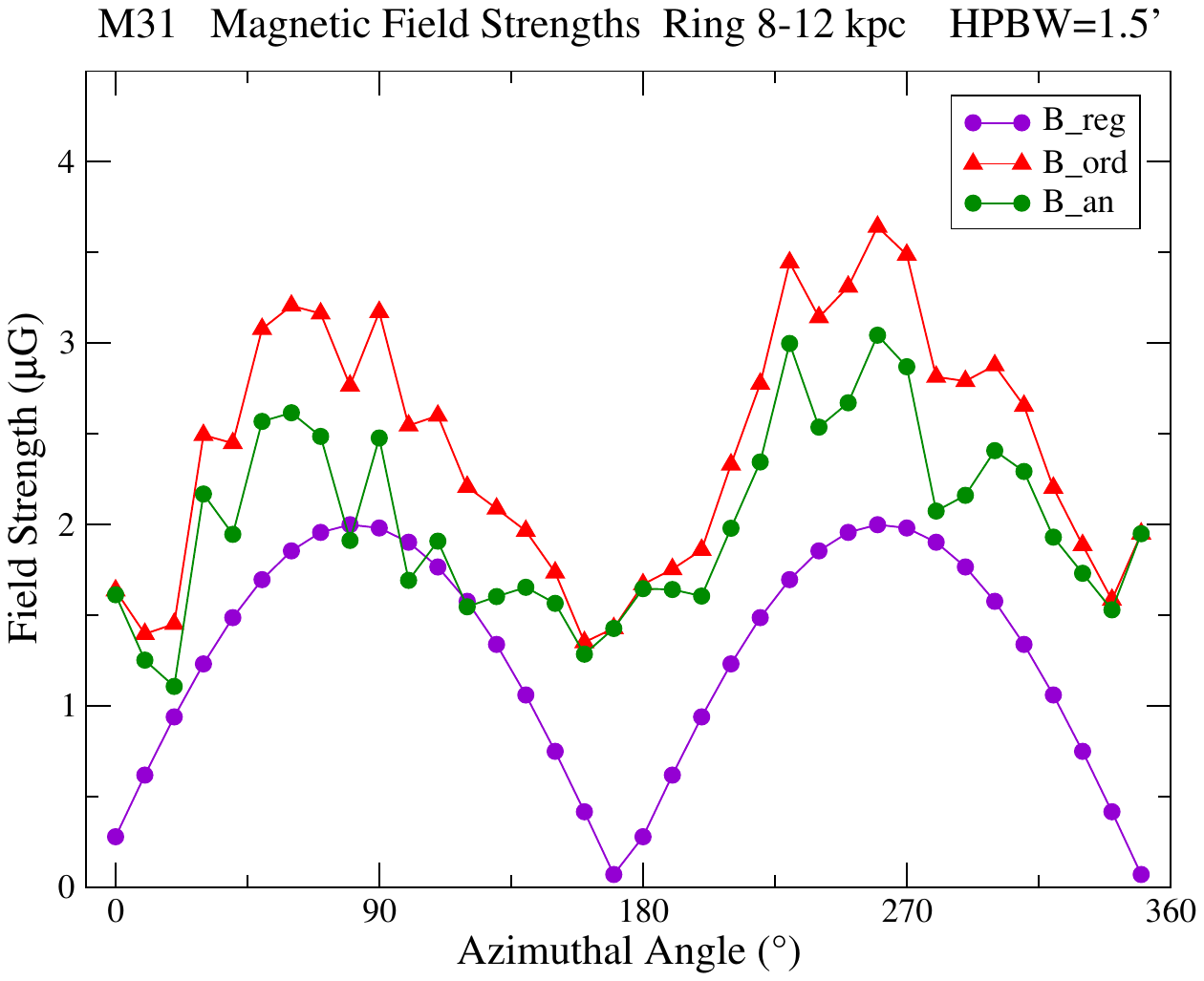}
\hfill
\caption{Variations in the strengths of the (plane-of-sky) ordered, axisymmetric (ASS) regular,
and anisotropic turbulent fields at $1\farcm5$ resolution with azimuthal angle, measured in the galaxy plane
in the radial ring 8--12\,kpc, using the spiral pitch angle of the ASS field of $-8\degr$
taken from the fit shown in Fig.~\ref{fig:rm3_6_azm}.}
\label{fig:Ban1}
\end{center}
\end{figure}

The ASS regular field in the sky plane, $B_\mathrm{reg,\perp}$, varies with azimuthal angle, $\phi$, as
\begin{equation}
B_\mathrm{reg,\perp} =  B_\mathrm{reg}\, | \, \sin\,(\phi - \xi_\mathrm{reg,\perp}) \, | \, .
\label{eq:Breg2}
\end{equation}

\noindent $B_\mathrm{reg}$ and $\xi_\mathrm{reg,\perp} \simeq  \xi_\mathrm{reg,\parallel}$ are taken
from Section~\ref{sec:Breg}. The anisotropic turbulent field $B_\mathrm{an,\perp}$ in the sky plane follows from
\begin{equation}
B_\mathrm{an,\perp} =  ( B_\mathrm{ord,\perp}^2 \, - \, B_\mathrm{reg,\perp}^2 )^{\,0.5} \, .
\label{eq:Ban}
\end{equation}

\begin{figure}[htbp]
\begin{center}
\includegraphics[width=0.95\columnwidth]{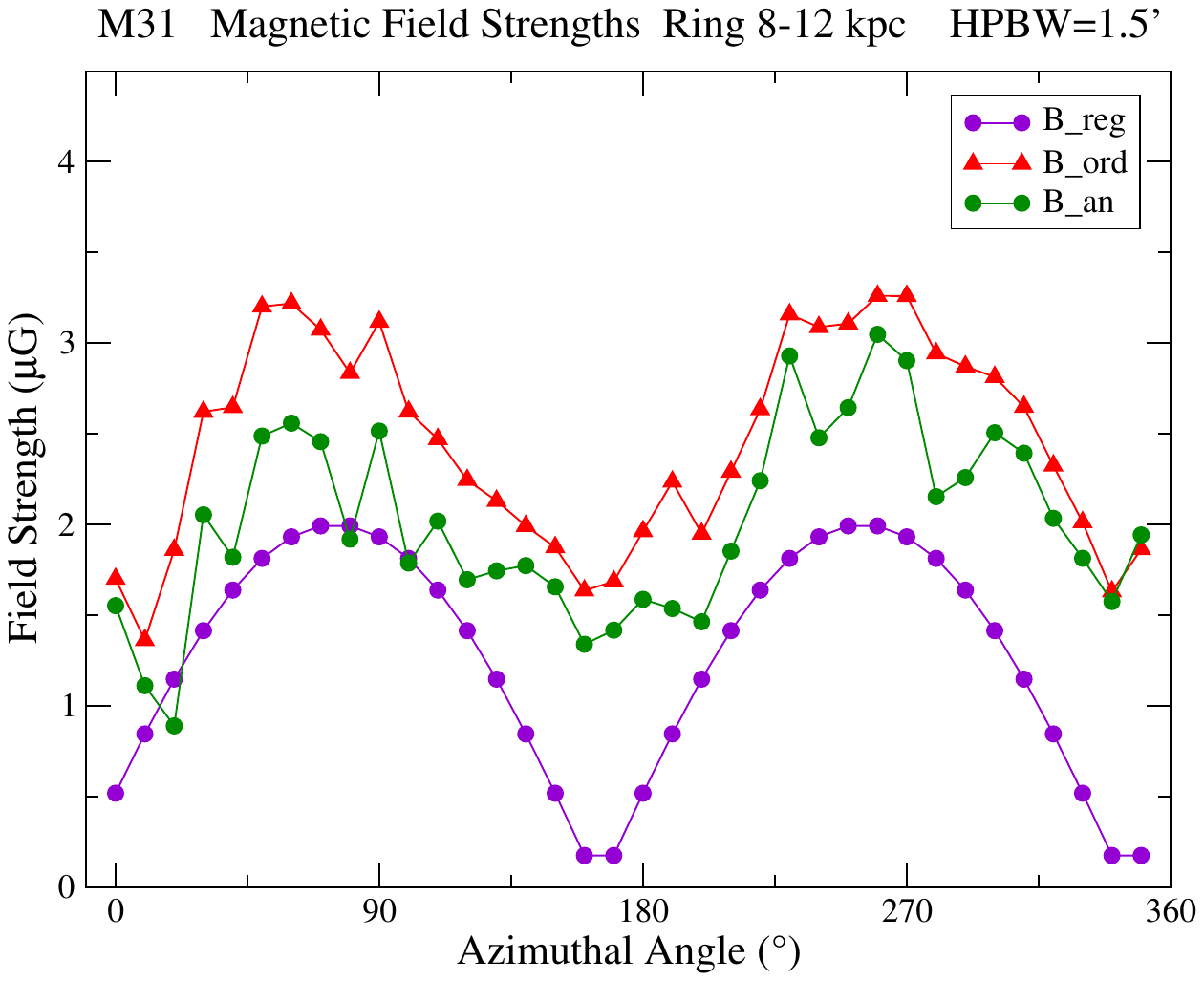}
\hfill
\caption{Variations in the strengths of the plane-of-sky ordered, axisymmetric (ASS) regular,
and anisotropic turbulent fields at $1\farcm5$ resolution with azimuthal angle, measured in the galaxy plane
in the radial ring 8--12\,kpc, assuming a spiral pitch angle of the ASS field of $-15\degr$.}
\label{fig:Ban2}
\end{center}
\end{figure}

The azimuthal variations in $B_\mathrm{ord,\perp}$, $B_\mathrm{reg,\perp}$, and $B_\mathrm{an,\perp}$
are shown in Figure~\ref{fig:Ban1}.
The anisotropic field strength shows a similar variation with azimuthal angle $\phi$
as that of $B_\mathrm{ord,\perp}$, though with a lower amplitude.
Averaging the ratio $\langle B_\mathrm{an,\perp} / B_\mathrm{ord,\perp} \rangle$ gives 0.84
with a low dispersion of 0.09.
Hence, the ordered field is dominated by the anisotropic turbulent field.
Around the minor axis (azimuthal angles $\phi\simeq90\degr$ and
$\simeq270\degr$) $B_\mathrm{reg,\perp}$ is largest and
almost reaches the level of $B_\mathrm{ord,\perp}$.
As both quantities are measured with independent methods, this demonstrates that the
equipartition assumption
is reasonable.

A phase shift between the variations in $B_\mathrm{an,\perp}$ and $B_\mathrm{reg,\perp}$
is obvious from Fig.~\ref{fig:Ban1}. One reason could be deviations from a constant spiral pitch angle of the
regular field, similar to what is observed for the gaseous spiral arms \citep{braun91,nieten06}.
The phase $\xi_\mathrm{reg,\parallel}$ of the RM variation in Fig.~\ref{fig:rm3_6_azm}
is mostly constrained by the regions around the minor axis, while the phase $\xi_\mathrm{reg,\perp}$
in Fig.~\ref{fig:Ban1}
is constrained by the regions around the major axis where the pitch angle
may be different.

Furthermore, $B_\mathrm{an}$ and $B_\mathrm{reg}$ may have different average spiral pitch angles,
as proposed by \citet{beck20}, based on the different phases the variation in RM and
of polarized intensity (their Table~6).
A steeper pitch angle of $\xi_\mathrm{reg,\perp}\simeq-15\degr$
brings the peaks of $B_\mathrm{an,\perp}$ and $B_\mathrm{reg,\perp}$ closer together
(Fig.~\ref{fig:Ban2}).

The azimuthal variation in RM in Fig.~\ref{fig:rm3_6_azm} shows deviations from
the fit, suggesting the existence of higher modes of the regular field with low
amplitudes, e.g. the bisymmetric (BSS) dynamo mode superimposed onto the ASS mode,
as suggested by \citet{sofue87a} and \citet{beck20}. We allowed the fit to include the higher modes
$\cos\,(2 \phi - \xi_2)$, $\cos\,(3 \phi - \xi_3)$, or $\cos\,(4 \phi - \xi_4)$ but none
improved the fit significantly.
The refined analysis by \citet[][submitted to MNRAS]{paul25} shows that several higher modes
contribute to the pattern of the regular field in M~31.

\begin{figure}[htbp]
\begin{center}
\includegraphics[width=0.95\columnwidth]{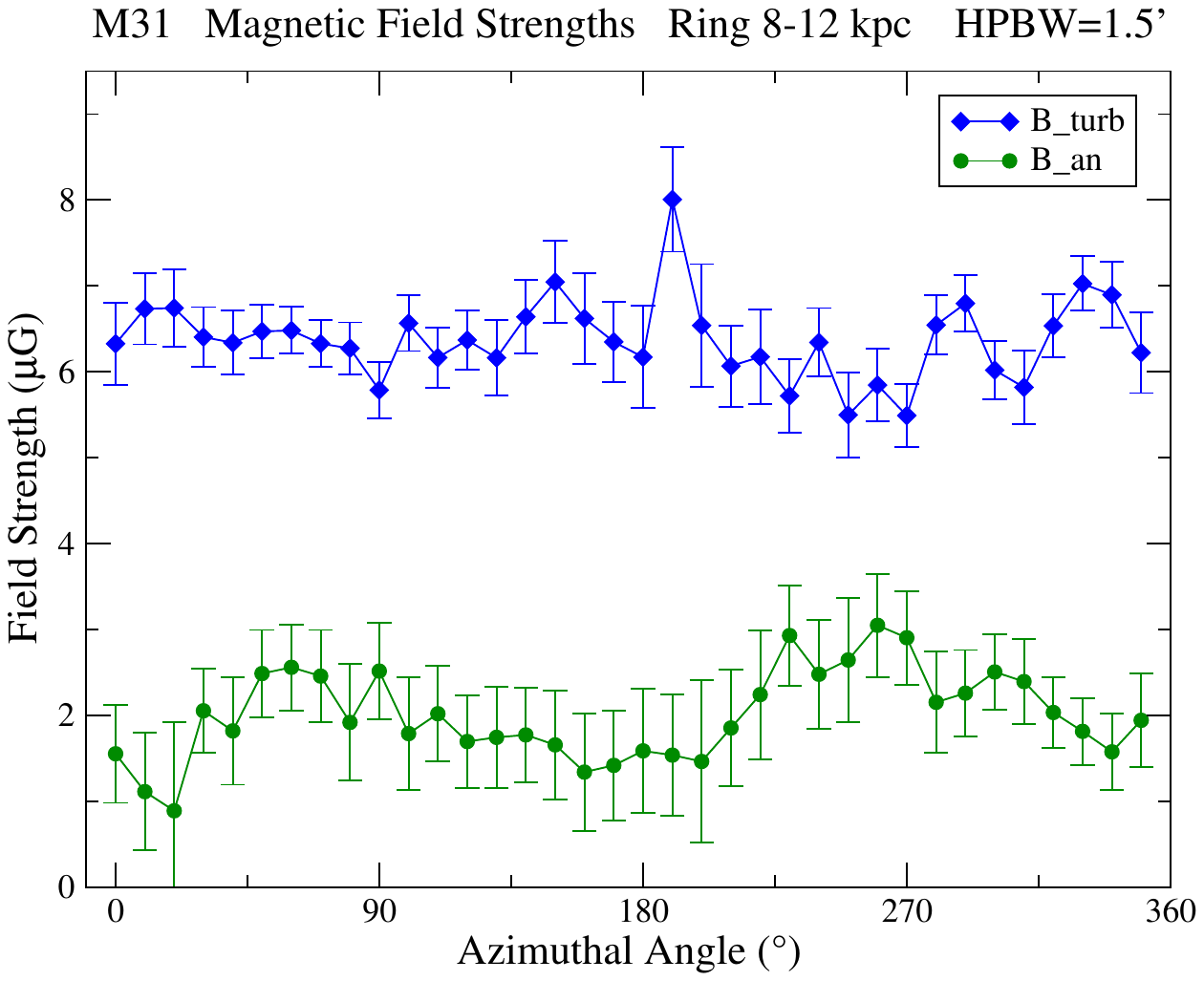}
\hfill
\caption{Variations in the isotropic turbulent field strength $B_\mathrm{turb}$
(Fig.~\ref{fig:Bmaps}, middle panel) and the strength of the (plane-of-sky) anisotropic turbulent
field $B_\mathrm{an,\perp}$ (taken from Fig.~\ref{fig:Ban2}) at $1\farcm5$ resolution
with azimuthal angle in the galaxy plane in the radial ring 8--12\,kpc.
The uncertainties in $B_\mathrm{turb}$ reflect the dispersion
of the values in the map of $B_\mathrm{turb}$ in each sector.
The uncertainties in $B_\mathrm{an,\perp}$ come from the dispersion of the values in the map of
$B_\mathrm{ord,\perp}$ (Fig.~\ref{fig:Bmaps}, bottom panel) in each sector
and the general uncertainty of $B_\mathrm{reg,\perp}$ as determined in Sect.~\ref{sec:Breg}.
}
\label{fig:Ban3}
\end{center}
\end{figure}

Fig.~\ref{fig:Ban3} shows the variations in the isotropic turbulent field strength, $B_\mathrm{turb}$,
and the anisotropic turbulent field strength, $B_\mathrm{an,\perp}$, with azimuthal angle
in the radial ring 8--12\,kpc. The average strength of $B_\mathrm{an,\perp}$
is $2.0\,\mu$G with a dispersion of $0.5\,\mu$G. Applying the same bias correction as for
$B_\mathrm{ord,\perp}$ (Sect.~\ref{sec:Bordeq}) yields $(2.7\pm0.7)\,\mu$G.
The average ratio $\langle B_\mathrm{an,\perp} / B_\mathrm{turb} \rangle$ from Fig.~\ref{fig:Ban3} is $0.32\pm0.10$
or $0.50\pm0.16$ after correction for bias.
The relatively small dispersion indicates that the generation of anisotropic turbulent fields
is related to isotropic turbulent fields. As large-scale density-wave shock fronts are weak or missing in M~31,
shearing by differential rotation is the probable origin of the anisotropy \citep{stepanov14,hollins17}.

\section{Radial variations in magnetic field strength}
\label{sec:radial}

Based on equipartition assumed to be valid on average within each
radial ring of 1\,kpc width in the galaxy plane, we computed the average strengths of the
total, turbulent, and ordered fields at $3\arcmin$ resolution as a function of radius.
We used the average values of the non-thermal intensities at \wave{20.5}, of the non-thermal spectral
index between \wave{6.2} and \wave{20.5} (Fig.~\ref{fig:radial}, middle panel), and of the
non-thermal degree of polarization at \wave{6.2} (Fig.~\ref{fig:radial}, bottom panel).
For the central region (0--3\,kpc radius), we used
a pathlength of $L_\mathrm{syn}=0.5$\,kpc and an inclination of $43\degr$ \citep{giessuebel14}.
For the radial range 3--20\,kpc, we used the same inclination of $75\degr$ and pathlength
of $L_\mathrm{syn}=2.5$\,kpc as for the emission torus (Section~\ref{sec:Btoteq}).
For the radial range 18--20\,kpc, where the uncertainties are large, we used a constant
non-thermal spectral index of 1.0 and a constant non-thermal degree of polarization of 17\%.

\begin{figure}[htbp]
\begin{center}
\includegraphics[width=0.95\columnwidth]{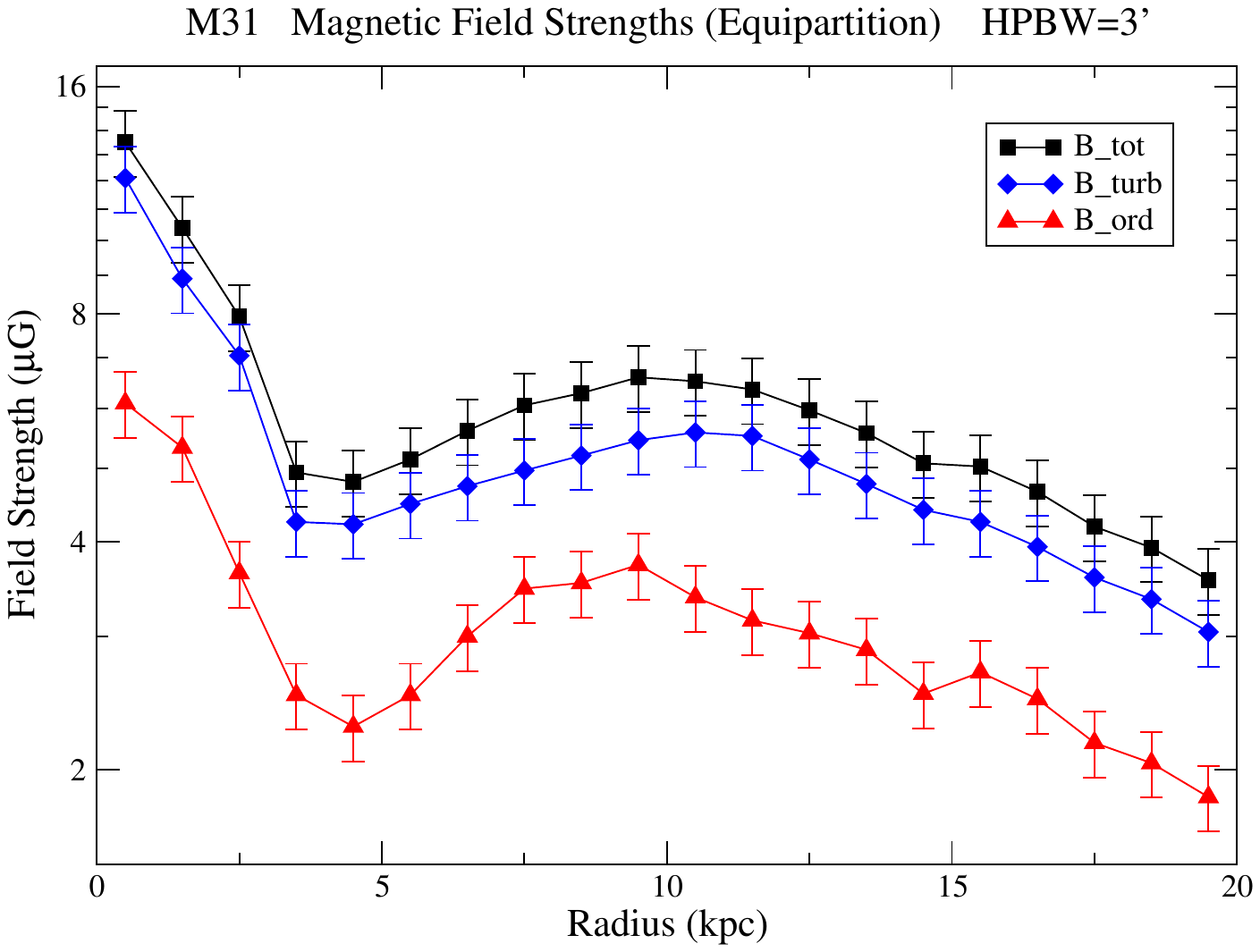}
\hfill
\caption{Average strengths (on a log scale) of the total, turbulent, and ordered fields
at $3\arcmin$ resolution in 1--kpc wide rings in the galaxy plane. The statistical
uncertainties are assumed to be 10\%. The systematical uncertainty of $\lesssim20\%$
(see Section~\ref{sec:errors}) affects all values by the same amount and is not included.}
\label{fig:B_radial}
\end{center}
\end{figure}

\begin{table*}
\begin{center}
  \caption{Radial variation in field strengths.
  }
  \label{tab:rm}
  \footnotesize
  \begin{tabular}{cccccccccccccc}
  \hline
  $\Delta R$ & $\mathrm{RM}_0$ & $\xi_\mathrm{reg,\parallel}$ & EM   &$\langle n_\mathrm{e} \rangle$ & $B_\mathrm{reg}$ & $B_\mathrm{turb}$ & $B_\mathrm{reg} /$  &
  $B_\mathrm{ord,\perp}$
  & $B_\mathrm{reg} /$ & $B_\mathrm{an,\perp}$ & $B_\mathrm{an,\perp} /$ & $B_\mathrm{an,\perp} /$ & $\delta$\\
  $[$kpc$]$  & [$\radm$] & [$\degr$] & [cm$^{-6}$\,pc]        &  [cm$^{-3}$]  & [$\mu$G]         & [$\mu$G]          & $B_\mathrm{turb}$   & [$\mu$G]
  & $B_\mathrm{ord,\perp}$ & [$\mu$G] & $B_\mathrm{turb}$ & $B_\mathrm{ord,\perp}$ \\
  \hline
  $7-8$   & $-93\pm13$  & $-4\pm5$ & 17.1  & 0.0284 & 1.96 & 4.97 & 0.39 & 3.47 & 0.56 & 2.86 & 0.58 & 0.82 & 1.15 \\
  $8-9$   & $-99\pm10$  & $-9\pm3$ & 18.3  & 0.0293 & 2.03 & 5.20 & 0.39 & 3.53 & 0.58 & 2.89 & 0.56 & 0.82 & 1.14 \\
  $9-10$  & $-120\pm9$  & $-7\pm3$ & 24.8  & 0.0342 & 2.11 & 5.45 & 0.39 & 3.73 & 0.57 & 3.08 & 0.57 & 0.83 & 1.15 \\
  $10-11$ & $-123\pm8$  & $-7\pm2$ & 28.8  & 0.0368 & 2.01 & 5.58 & 0.36 & 3.38 & 0.59 & 2.72 & 0.49 & 0.80 & 1.11 \\
  $11-12$ & $-129\pm11$ & $-5\pm3$ & 23.1  & 0.0329 & 2.35 & 5.52 & 0.43 & 3.15 & 0.75 & 2.10 & 0.38 & 0.67 & 1.07 \\
  \hline
  \end{tabular}
\end{center}
\tablefoot{
$\Delta R$: radial range in the galaxy plane, $\mathrm{RM}_0$: RM amplitude,
  $\xi_\mathrm{reg,\parallel}$: phase of the RM variation with azimuthal angle (from \citet{beck20}),
  EM: emission measure,
  $\langle n_\mathrm{e} \rangle$: average thermal density,
  $B_\mathrm{reg}$: strength of the regular field,
  $B_\mathrm{turb}$: strength of the isotropic turbulent field,
  $B_\mathrm{ord,\perp}$: strength of the ordered field in the sky plane,
  $B_\mathrm{an,\perp}$: strength of the anisotropic turbulent field in the sky plane,
  $\delta$: degree of anisotropy ($\delta=({1+B_\mathrm{an}^2/B_\mathrm{turb}^2})^{\,0.5}$),
  all at $3\arcmin$ resolution.
}
\end{table*}
\normalsize

\begin{figure}[htbp]
\begin{center}
\includegraphics[width=0.95\columnwidth]{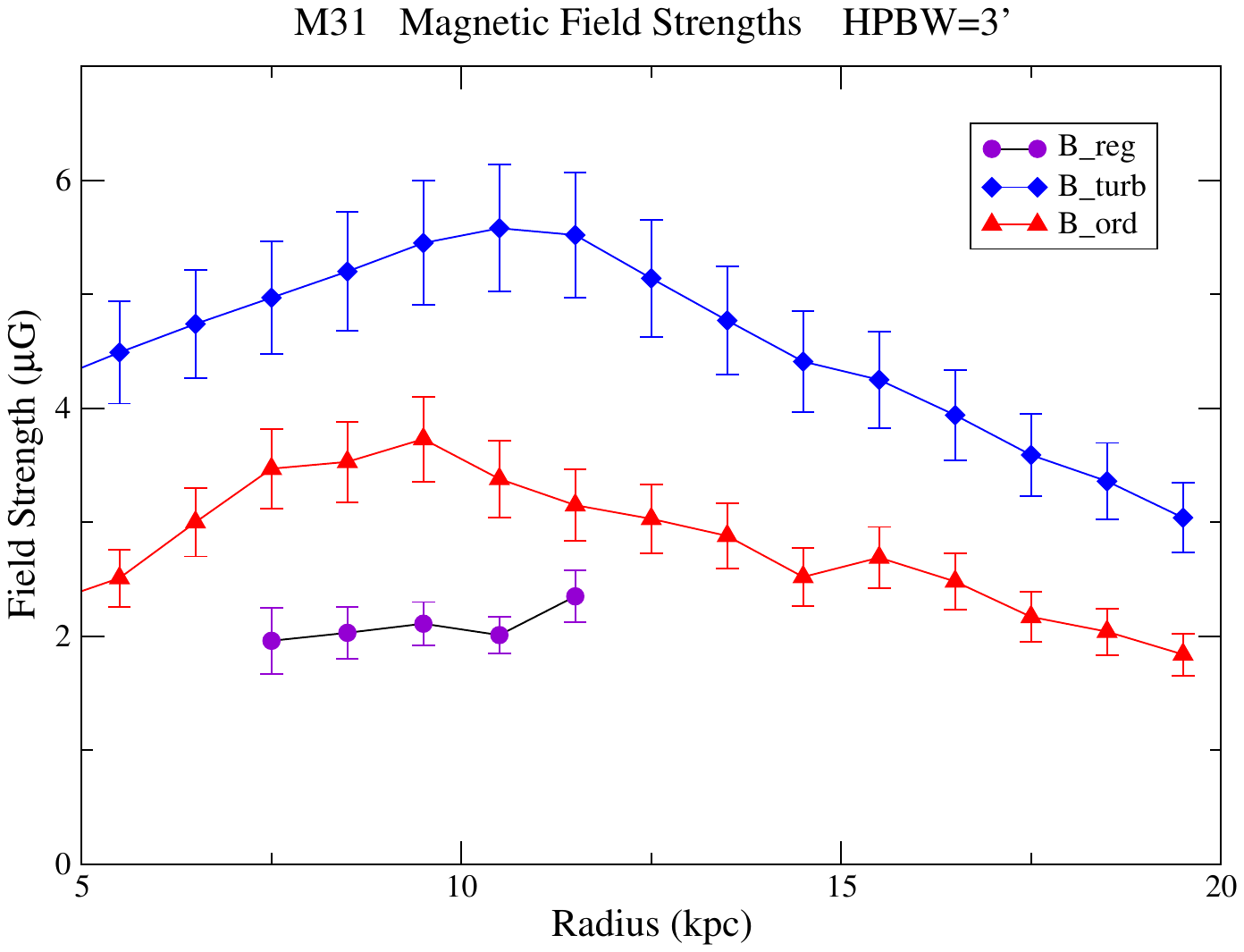}
\hfill
\caption{Average strengths (on a linear scale) of the turbulent and ordered fields
(from Fig.~\ref{fig:B_radial}) and of the regular field (Table~\ref{tab:rm})
at $3\arcmin$ resolution in 1--kpc wide rings in the galaxy plane.
The uncertainties in $B_\mathrm{reg}$ are estimated from the statistical uncertainties
in RM and assuming a 5\% uncertainty in $\langle n_\mathrm{e} \rangle$. The systematic uncertainties
in $L_\mathrm{th}$ and $f_\mathrm{th}$ of about 18\% and 25\%, respectively,
affect all $B_\mathrm{reg}$ values by the same amount and are not included.}
\label{fig:B_radial2}
\end{center}
\end{figure}

The result is shown in Figure~\ref{fig:B_radial}. The total and ordered fields
peak at about 10\,kpc radius, the turbulent field peaks at about 11\,kpc
radius, always followed by a slow decrease with increasing radius.
Fits to the averages of total, turbulent, and ordered fields in the radial range
11--20\,kpc by exponential functions $B\propto \exp(-R/R_0)$ give similar scale lengths of
$R_0=(14.1\pm0.6)$\,kpc, $R_0=(13.8\pm0.4)$\,kpc, and $R_0=(15.3\pm1.4)$\,kpc, respectively.
Compared with the exponential scale length of the total non-thermal emission at
\wave{20.5} at $3\arcmin$ resolution of $(3.1\pm0.1)$\,kpc (Section~\ref{sec:AME}),
the scale length of the total magnetic field is larger by a factor of $4.5\pm0.2$.
This is consistent with the expectation from the assumption of energy
equipartition for which a factor of $(3+\alpha_\mathrm{n})\simeq4$ is expected.

Exponential scale lengths of the total equipartition field were estimated in five
spiral galaxies by \citet{basu13}, with values ranging between 5\,kpc and 18\,kpc.
\citet{berkhuijsen16} found a scale length in the outer disc of the spiral galaxy
M~101 of about 20\,kpc for the total and turbulent fields.

Profiles of the total magnetic field strength were determined for 198 analogues
of the Milky Way and M~31 in the TNG50 simulation \citep[Figure~19 in][]{pillepich24}.
The average strength is 5\,$\mu$G at 10\,kpc radius and 3\,$\mu$G at 20\,kpc radius,
similar to our observations (Fig.~\ref{fig:B_radial}).

The radial variation in the strength $B_\mathrm{reg}$ of the regular field was
obtained from the fit results of the sinusoidal azimuthal variation
of RM between \wave{3.6} and \wave{6.2} at $3\arcmin$ resolution
in five rings between 7\,kpc and 12\,kpc \citep{beck20}.
To estimate the densities of the diffuse warm ionized gas $\langle n_\mathrm{e} \rangle$ in these
rings with help of Equations (\ref{eq:EM}) and (\ref{eq:ne}), we measured the thermal intensities
$\langle I_\mathrm{th} \rangle$ in the same five rings
from the map shown in Fig.~\ref{fig:cm3th}, smoothed to $3\arcmin$ resolution. We assumed the
same constant pathlength through the warm ionized gas of $L_\mathrm{th}=4.25$\,kpc and constant volume
filling factor $f_\mathrm{th}=0.2$ (Section~\ref{sec:Breg}).

The strengths of the regular field, $B_\mathrm{reg}$, in the five rings follow from
Eq.~(\ref{eq:Breg3}), are listed in Table~\ref{tab:rm}, and are shown in Figure~\ref{fig:B_radial2}.
$B_\mathrm{reg}$ is on average smaller by about 40\% than the equipartition strength of
$B_\mathrm{ord,\perp}$ (containing regular and anisotropic turbulent fields).
As $B_\mathrm{reg}$ and $B_\mathrm{ord,\perp}$ were measured with independent methods,
this gives
indication that the data are consistent with the equipartition assumption.

Table~\ref{tab:rm} gives the strengths of the isotropic turbulent field $B_\mathrm{turb}$,
the ordered field $B_\mathrm{ord,\perp}$, and the anisotropic turbulent field
$B_\mathrm{an,\perp}$ in the sky plane
(derived from $B_\mathrm{an,\perp}^2 = B_\mathrm{ord,\perp}^2 - B_\mathrm{reg}^2$).
The ratio $B_\mathrm{reg}/B_\mathrm{turb}\simeq0.39$ is almost constant over all five rings.
The ratio $B_\mathrm{an,\perp}/B_\mathrm{turb}\simeq0.57$
and the degree of anisotropy $\delta\simeq1.15$ are constant until 10\,kpc radius
and then decrease.
Similarly, the ratio $B_\mathrm{an,\perp}/B_\mathrm{ord,\perp}\simeq0.82$
is constant until 10\,kpc radius (similar to the average ratio in Fig.~\ref{fig:Ban2})
and then decreases.

$B_\mathrm{reg}$ remains about constant in the radial range 7--12\,kpc,
while $B_\mathrm{ord,\perp}$ decreases (Fig.~\ref{fig:B_radial2}). If confirmed by
future data at larger radii,
possible reasons for this divergence are:\\
(1) The anisotropic turbulent field, $B_\mathrm{an,\perp}$, decreases radially faster than the
regular field $B_\mathrm{reg}$ and gradually contributes less to the ordered field
(see last column of Table~\ref{tab:rm}). $B_\mathrm{an,\perp}$ decreases
because the shear rate
\footnote{For a flat rotation curve $\mathrm{v_{rot}}$\,=\,const, the shear rate is\\
$R\,(d\Omega/dR) = - \mathrm{v_{rot}}/R$.}
decreases with radius together with the gradient of angular velocity $\Omega$,
while the large-scale dynamo generating $B_\mathrm{reg}$ remains efficient until large radii.\\
(2) If the disc of warm thermal gas flares beyond about 10\,kpc radius, as proposed by \citet{fletcher04},
the pathlength $L_\mathrm{th}$ increases with radius and $B_\mathrm{ reg}$ decreases with
$L_\mathrm{th}^{-0.5}$ (Eq.~\ref{eq:Breg3}).

\section{Energy densities}
\label{sec:energy}

The dynamical importance of magnetic fields can be estimated by comparing the magnetic energy density
with the thermal energy density of the diffuse warm ionized gas and the kinetic energy density of the turbulent
neutral gas. We restricted our estimates to the emission torus at 8--12\,kpc radius.

\begin{itemize}

\item The average equipartition strength of the total field in the emission torus (radial range 8--12\,kpc)
of 6.3\,$\mu$G (Sect.~\ref{sec:Btoteq}) corresponds to a magnetic energy density of
$\epsilon_\mathrm{mag} \simeq 1.58\cdot10^{-12}$\,erg\,cm$^{-3}$.
For a systematic uncertainty in $B$ of $\lesssim 20\%$, the systematic uncertainty in
$\epsilon_\mathrm{mag}$ is $\lesssim 40\%$.

\item To compute the kinetic energy density of the turbulent neutral gas, $\epsilon_\mathrm{kin}$,
the mass density, $\langle \rho \rangle$, and the turbulent velocity, v$_\mathrm{turb}$, need to be known.
We computed a map of the column densities of the neutral gas from the HI and CO data
\citep{braun91,nieten06} and measured $N_\mathrm{gas}\simeq 2.6 \cdot 10^{21}$\,cm$^{-2}$.
The scale height of HI gas in the emission torus of $H_\mathrm{gas} \simeq 350$\,pc \citep{braun91},
similar to the synchrotron scale height at \wave{20.5}
(Sect.~\ref{sec:Btoteq}), yields a pathlength of
$L_\mathrm{gas}\simeq2.7$\,kpc and hence a volume density of the HI gas of $\langle n \rangle \,= 0.31$\,cm$^{-3}$
and a mass density of $\langle \rho \rangle \,= 5.2 \cdot 10^{-25}$\,g\,cm$^{-3}$.
Allowing for a helium contribution of $\simeq30\%$ to the mass yields
$\langle \rho \rangle \,= 6.8 \cdot 10^{-25}$\,g\,cm$^{-3}$.
\citet{caldu13} measured a typical turbulent velocity of v$_\mathrm{turb}=12$\,km/s for nearby spiral galaxies
from spectral line observations that refers to one dimension. For three dimensions, we get
$\epsilon_\mathrm{kin}=1.5 \, \langle \rho \rangle \, \mathrm{v}_\mathrm{turb}^2 \simeq 1.4\cdot10^{-12}$\,erg\,cm$^{-3}$.
This number could slightly increase if the spin temperature of 20--60\,K assumed by \citet{braun09}
is too low, as indicated by HI data from the Milky Way \citep{dickey88,basu22}, so that the HI column
density could be underestimated.\\
In conclusion, the kinetic and magnetic energy densities are similar.

\item The thermal energy density of the diffuse warm ionized medium (WIM) is
$\epsilon_\mathrm{th,WIM}=2.1 \, n_\mathrm{e} \, k \, T$ \citep{ferriere01},
where $n_\mathrm{e} = \langle n_\mathrm{e} \rangle /f_\mathrm{th}$ is the local electron density
within a cloud.
\footnote{We note that $\langle n_\mathrm{e} \rangle$ instead of $n_\mathrm{e}$ was used
in previous estimates of $\epsilon_\mathrm{th}$ in several spiral galaxies \citep{beck07,beck15d,taba08},
resulting in too small values of $\beta$.}
With $\langle n_\mathrm{e} \rangle \,= (0.034\pm0.005)$\,cm$^{-3}$, $T=(7000\pm1000)$\,K, and
$f_\mathrm{th}=0.20\pm0.05$ (Sect.~\ref{sec:Breg}), we get
$\epsilon_\mathrm{th,WIM}\simeq (3.5\pm1.1) \cdot 10^{-13}$\,erg\,cm$^{-3}$.
The plasma--$\beta$ parameter is $\beta=\epsilon_\mathrm{th,WIM}/\epsilon_\mathrm{mag}\simeq 0.22$
with an uncertainty of $\lesssim 0.11$.
The average magnetic energy density in the emission torus of M~31 is 4.6\,times (with an uncertainty
of $\lesssim 2.3$) larger than the thermal energy density of the warm ionized gas.

The total magnetic field strength in the local ISM of the Milky Way
of $(6.1\pm0.5)$\,$\mu$G \citep{han04} and the local thermal density of the warm ionized gas of
$\langle n_\mathrm{e} \rangle \,= (0.025\pm0.005)$\,cm$^{-3}$ \citep[Table~1 in][]{gaensler08},
with the same $T$ and $f_\mathrm{th}$ as in M~31, gives $\beta\simeq0.17\pm0.07$, similar to our result for M~31.
As $\epsilon_\mathrm{kin} > \epsilon_\mathrm{th,WIM}$, turbulence appears to be supersonic in the warm ionized
gas in M~31 as well as in the solar neighbourhood.

\item The thermal energy density of the diffuse warm neutral medium (WNM) is
$\epsilon_\mathrm{th,WNM}=1.1 \, \langle n \rangle \, k \, T$ \citep{ferriere01}.
$\langle n \rangle \,= 0.31$\,cm$^{-3}$ (see above), plus 10\% for helium,
and $T=5000$\,K \citep{ferriere20} give $\epsilon_\mathrm{th,WNM}\simeq 2.6 \cdot 10^{-13}$\,erg\,cm$^{-3}$,
similar to $\epsilon_\mathrm{th,WIM}$.

\end{itemize}

\section{Discussion and conclusions}
\label{sec:conclusions}

In this paper, separate maps of the strengths of the various components of interstellar
magnetic fields in M~31 (i.e. isotropic turbulent, anisotropic turbulent, ordered, and regular) are
computed for the first time, based on physically reasonable assumptions, such as that of
energy density equipartition between total magnetic fields and total CRs,
and with proper estimates of the uncertainties. Our results are of fundamental importance
for comparisons with models of magnetic field origin.

The equipartition assumption has been a matter of intense discussion.
\citet{stepanov14} and \citet{seta19} argued that equipartition is not valid on scales smaller than the
CR propagation scale of a few 100\,pc.
\citet{tharakkal23a} showed from numerical modelling that equipartition is not valid on spatial scales
smaller than the correlation scale of magnetic fields; CRs spend more time in magnetic traps
where the magnetic field is weaker.
\citet{ponnada22,ponnada24} modelled synchrotron intensities in spiral galaxies and concluded that
equipartition is roughly valid on scales of $>1$\,kpc but not valid on smaller scales due to
the small volume filling factor of the emission.
In contrast, the synchrotron intensities predicted from a MHD simulation of a Milky Way-like galaxy,
assuming equipartition, are at least an order of magnitude lower than the ones observed \citep{dacunha25}.
These authors suggested that the equipartition field strengths inferred from observations are typically
a factor of 2--3 too high and may even be overestimated by two dex in low-intensity inter-arm regions.

\citet[][Section 4.3.2]{ruszkowski23} argued that there is no physical mechanism that ensures
equipartition, found that in starburst and dwarf galaxies equipartition is invalid,
and summarized that energy equipartition needs time to develop and may not be applicable on
small time and spatial scales.

The total equipartition field in the ISM of the local Milky Way of $\simeq6$\,$\mu$G
\citep[Fig.~5 in][]{beck96} was confirmed by
the in situ measurements of the two Voyager spacecrafts just outside of the solar system
of $(4.8\pm0.4)$\,$\mu$G and $(6.8\pm0.3)$\,$\mu$G, respectively \citep{brandt23}.\\

Our main results are :
\begin{itemize}

\item The strength $B_\mathrm{reg}$ of the regular field is on average about 40\% smaller than
the strength $B_\mathrm{ord,\perp}$ of the ordered field (containing regular and anisotropic
turbulent fields). As those two quantities were measured
with independent methods (i.e. Faraday rotation and polarized synchrotron emission),
this is consistent with
the assumption of energy density equipartition between total magnetic fields and total CRs.

\item A comparison of our maps of synchrotron emission at \wave{3.6} and \wave{20.5} yields
an estimate of the diffusion length of CREs emitting at \wave{3.6} of
$\lesssim 0.34$\,kpc in the sky plane, setting a lower limit for the validity of the
equipartition assumption.

\item Fluctuations in field strength are used to estimate a volume filling factor of
the total magnetic field in the diffuse ISM of $\simeq0.96$, which is much larger than in the MHD model of the small-scale
dynamo by \citet{ponnada24}, while models of the mean-field dynamo including
tangling of the regular field by turbulence produce a turbulent field that fills the
volume and is comparable in strength to the regular field \citep[Section~13.3 in][]{shukurov21}.

\item The magnetic field energy $\epsilon_\mathrm{mag}$ is a primary dynamical agent
in the ISM of M~31 ($\epsilon_\mathrm{mag}/\epsilon_\mathrm{th}\simeq5$ or
plasma--$\beta \simeq 0.2$ for the warm ionized gas) and
reaches the level of kinetic energy ($\epsilon_\mathrm{mag} \simeq \epsilon_\mathrm{kin}$).
For the spiral galaxy M~33,
\citet{taba08} found $\epsilon_\mathrm{mag} \simeq \epsilon_\mathrm{kin}$ in the whole disc,
as well as for the inner discs of the massive spiral galaxies NGC~6946 \citep{beck07} and IC~342
\citep{beck15d}, while $\epsilon_\mathrm{mag}$ dominates in the outer discs of NGC~6946 and IC~342.

The similarity between the energy densities of the total magnetic field and the turbulent gas
motions is in conflict with numerical
MHD simulations of the small-scale dynamo
\citep[e.g.][]{federrath14,rieder17,seta22,gent23}, which predict
a ratio between magnetic and kinetic energy densities of only a few percent.
Higher saturation levels can be reached in massive galaxies with fast rotation \citep{pakmor24}.

Present-day numerical MHD simulations are still insufficient to model the magnetic ISM in a realistic way.
Possible reasons could be the limited spatial resolution, which prevents the small-scale dynamo from
developing a strong turbulent field. The warm ionized ISM has a Reynolds number of about $5\cdot10^7$ and a magnetic
Prandtl number of about $10^{11}$ \citep[Table~2 in][]{ferriere20}, many orders of magnitude larger
than what is achieved in simulations. The lack of resolution in low-density regions can globally
distort the magnetic field \citep[Section~13.14.2 in][]{shukurov21}.

Furthermore, none of the above simulations included the mean-field dynamo.
Modelling the simultaneous action of the small-scale and mean-field dynamos, \citet{gent24}
showed that tangling of the regular (mean) field is able to amplify small-scale fields beyond the
saturation level, possibly up to the equipartition level with kinetic energy.

\item Thanks to the high inclination of M~31, RMs are strong and
allowed us to measure the mean strength of the regular field of $(2\pm0.5)\,\mu$G.
This value is consistent with the results of the large-scale dynamo model for massive galaxies
presented by \citet{rodrigues19}.

\item The prominent sinusoidal RM variation with azimuthal angle in the emission torus of M~31
enabled us to construct a 3D model of the regular field.
An exceptionally efficient large-scale dynamo operates in the emission torus of M~31 and generates
a regular field with a dominating basic mode characterized by an ASS pattern.
In all other spiral galaxies observed so far, the regular field reveals a spectrum of dynamo modes
\citep[Table~5 in][]{beck19}, so that more effort is needed to construct a 3D model.

\item The ratio between the strengths of regular and isotropic turbulent fields
of $\simeq0.39$ is almost constant with the azimuth in the 8--12\,kpc ring
as well as almost constant with the radius over the radial range 7--12\,kpc.
A semi-analytical simulation of the simultaneous operation of the small-scale and the mean-field dynamos
by \citet{bhat16} gave a ratio of $\simeq0.2$ for their largest (though still too small) Reynolds numbers.
Improved MHD models are being developed \citep{gent24} and will profit from the result obtained
in this paper.

\item The ratio between the strengths of the anisotropic and isotropic turbulent
fields of $\simeq0.57$ is almost constant from 7 to 10\,kpc and
decreases towards larger radii.
This indicates that shearing of isotropic turbulent fields by differential rotation
is an important mechanism to generate anisotropic turbulent fields, as has been proposed by \citet{hollins17}.

\end{itemize}

New insights into the magnetic field of M~31 can be expected from deep observations of a large number
of polarized background sources and their RMs, expanding the work by \citet{han98},
to investigate the detailed field structure. Rotation measure data of the diffuse
synchrotron emission of M~31 with improved sensitivity and angular resolution are required for
a better understanding of the origin of the various field components.
While the southern locations of MeerKAT and the Square Kilometre Array (SKA,
under construction) hamper observations of M~31, the Jansky Very Large Array
(JVLA) is a suitable instrument for such investigations.
\\

\noindent This is the last in a series of more than 30 refereed papers on M~31 published by us since 1973.
\\

\begin{acknowledgements}

We wish to thank many colleagues for inspiring discussions, namely Aritra Basu, Luke Chamandy,
Katia Ferri\`ere, Andrew Fletcher, Ren\'e Gie\ss\"ubel, Jürgen Kerp, Sui Ann Mao, Rüdiger Pakmor, Wolfgang Reich, Amit Seta,
Anvar Shukurov, Dmitry Sokoloff, and Fatemeh Tabatabaei.
We thank the anonymous referee for many valuable suggestions.

\end{acknowledgements}

\bibliographystyle{aa} 
\bibliography{m31}

\begin{appendix}

\section{Faraday depolarization}
\label{sec:dp}

The results from this paper can be used for a consistency check with help of another
observable, internal Faraday dispersion, which is caused by turbulent magnetic fields embedded in ionized gas
\citep{sokoloff98,arshakian11}, and dominates in spiral galaxies \citep{williams24}.

Additional depolarization could be caused by differential Faraday rotation by regular fields and depends on RM
\citep{sokoloff98,arshakian11}. Although RMs in M~31 are sufficiently large to cause significant
depolarization at \wave{20.5}, internal Faraday dispersion reduces the observable layer of polarized emission
and reduces the effective RM.

For an emitting and Faraday-rotating region, internal Faraday dispersion reduces the degree of
polarization $p_\mathrm{n,0}$ of synchrotron emission by the factor $\mathrm{DP}_\lambda$:
\begin{equation}
\mathrm{DP}_\lambda \, = p_\mathrm{n} \, / \, p_\mathrm{n,0} \, = \, (1 - \mathrm{exp}(-S)) / S \, ,
\label{eq:dp1}
\end{equation}

\noindent where $S = 2 \sigma_\mathrm{RM}^2\,\lambda^4$ at short wavelengths where Faraday effects are
small or moderate ($S\leq1$), as is the case for M~31 at \wave{3.6} and \wave{6.2}.
$\sigma_\mathrm{RM}$ is the dispersion in the RM. In a random-walk approach,
$\sigma_\mathrm{RM} = 0.812 \, B_\mathrm{turb,\parallel} \,\, n_\mathrm{e} \, d \, N_\parallel^{1/2}$
\citep{beck07}, where $B_\mathrm{turb,\parallel} = \sqrt{1/3}\,\,B_\mathrm{turb}$ is the
strength of the isotropic turbulent field and $n_\mathrm{e}$ the electron density within
each turbulent cell of size $d$.
$N_\parallel = L  \, f / d$ is the number of cells along the line of sight, L, with a volume filling factor f.
The volume-averaged electron density along the line of sight is
$\langle n_\mathrm{e} \rangle = n_\mathrm{e}\cdot f_\mathrm{th}$, so that we get
\begin{equation}
\sigma_\mathrm{RM} = 0.812 \, B_\mathrm{turb,\parallel} \, \langle n_\mathrm{e} \rangle \, \sqrt{L_\mathrm{syn} \, d / f_\mathrm{th}} \, .
\label{eq:dp2}
\end{equation}

$\sigma_\mathrm{RM}$ was computed as a function of radius in the plane of M~31 from $B_\mathrm{turb}$ derived in
Section~\ref{sec:radial} and $\langle n_\mathrm{e} \rangle$ from the emission measure EM, as described in
Section~\ref{sec:Bordeq}.
We used $d\simeq50$\,pc from \citet{fletcher11} and $f_\mathrm{th}=0.2$ from Sect.~\ref{sec:Breg}.
The pathlength was assumed to be the same as that through the region of
synchrotron emission, $L_\mathrm{syn}=2.5\,$kpc, where the turbulent field is located.
The depolarization factor at \wave{6.2}, $\mathrm{DP}_6$, was computed from Eq.~(\ref{eq:dp1}).
However, application of Eq.~(\ref{eq:dp1}) to \wave{20.5} would lead to DP factors much smaller
than observed.

Internal Faraday dispersion causes the (spatial) correlation length of the polarized emission to
decrease with increasing wavelength. Beyond the wavelength where the correlation length drops below
the size of the turbulent cells, no extended polarized emission is visible anymore, as is the case for M~31
at \wave{20.5} \citep{beck98}.
\citet{tribble91} developed a model for depolarization by an external Faraday screen at long wavelengths.
A model of depolarization by internal Faraday dispersion at long wavelengths ($S>1$) is still lacking.
Assuming that the results by \citet{tribble91} can be applied to our data at \wave{20.5}, the depolarization factor is
\begin{equation}
\mathrm{DP}_{20} \, = \, 1\,/ \,(2\,\sigma_\mathrm{RM}\,\lambda^2)\, .
\label{eq:dp2}
\end{equation}

\begin{figure}[htbp]
\begin{center}
\includegraphics[width=0.95\columnwidth]{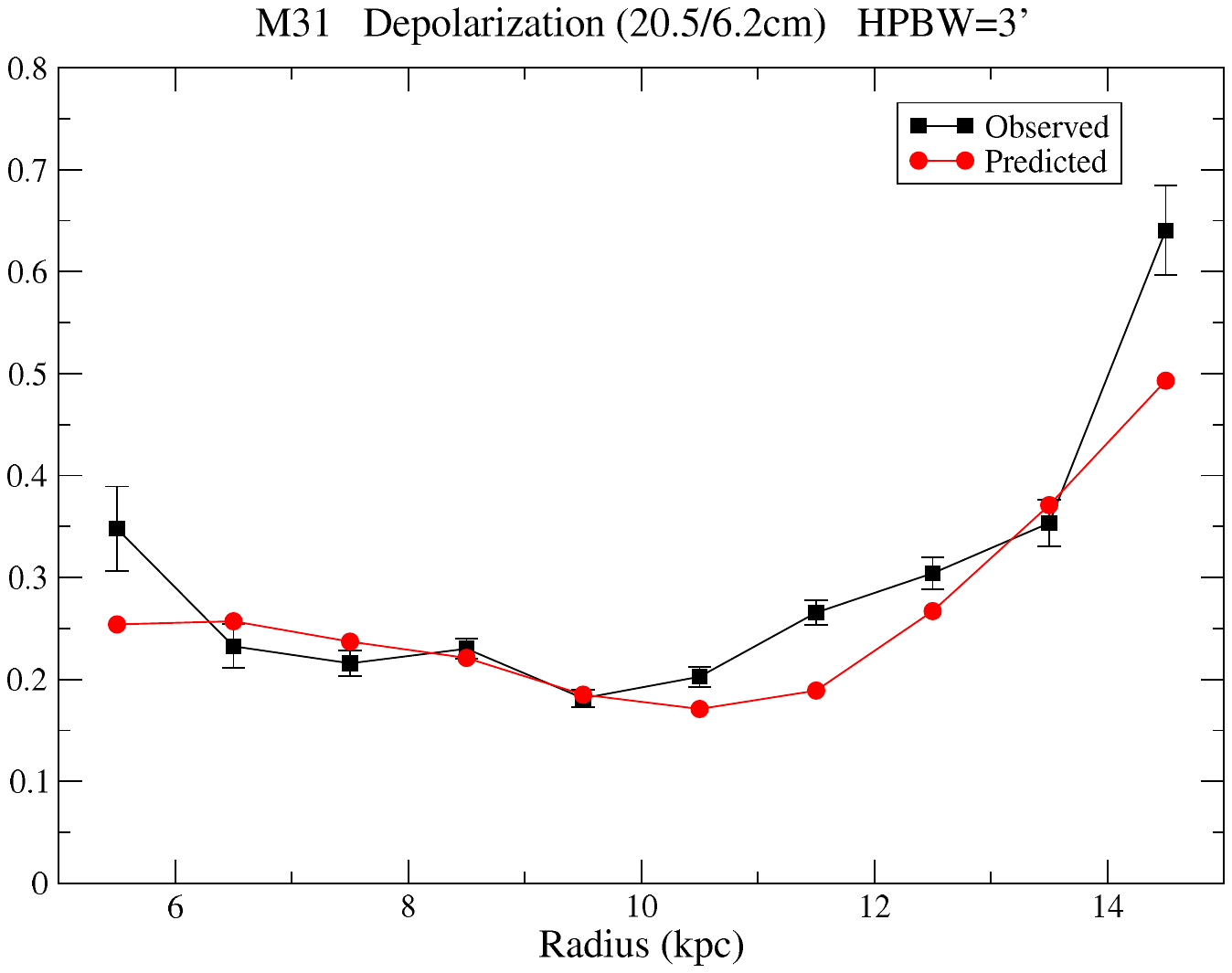}
\hfill
\caption{Observed depolarization $\mathrm{DP}=p_\mathrm{n,20}/p_\mathrm{n,6}$ between \wave{20.5} and \wave{6.2}
in 1--kpc wide rings in the galaxy plane at $3\arcmin$ resolution and predicted depolarization
$\mathrm{DP}=\mathrm{DP}_{20}/\mathrm{DP}_6$ by internal Faraday dispersion, using the observed variation in isotropic turbulent fields
$B_\mathrm{turb}$ and thermal electron densities $\langle n_\mathrm{e} \rangle$. The statistical uncertainty
in $B_\mathrm{turb}$ of $\simeq10\%$ leads to an uncertainty of the predicted DP of $\simeq7\%$.
The systematical uncertainty in $\langle n_\mathrm{e} \rangle$ of $\simeq20\%$ causes a systematical uncertainty
in DP of $\simeq14\%$.
}
\label{fig:dp_radial}
\end{center}
\end{figure}

Figure~\ref{fig:dp_radial} shows the observed depolarization $\mathrm{DP}=p_\mathrm{n,20}/p_\mathrm{n,6}$
compared to the ratio of depolarization factors $\mathrm{DP}=\mathrm{DP}_{20}/\mathrm{DP}_6$ as predicted
by our model (Table~\ref{tab:dp}). The agreement is excellent. This gives support that our magnetic
field measurements obtained from the equipartition assumption as well as Eq.~\ref{eq:dp2} are reasonable.

A model for internal Faraday depolarization at long wavelengths is needed for the analysis of
radio polarization data from the LOFAR and SKA-Low telescopes.

\begin{table}[h!]
\begin{center}
  \caption{Radial variation in Faraday depolarization.
  }
  \label{tab:dp}
  \footnotesize
  \begin{tabular}{ccccccc}
  \hline
    $\Delta R$  & $B_\mathrm{turb}$ & $\langle n_\mathrm{e} \rangle$ & $\sigma_\mathrm{RM}$ & $\mathrm{DP}_6$ & $\mathrm{DP}_{20}$ & $\mathrm{DP}$ \\
  $[$kpc$]$   & [$\mu$G]          & [cm$^{-3}$]      &  [rad\,m$^{-2}$]        & & &  \\
  \hline
  $5-6$   & 4.49 & 0.0293 & 48.8 & 0.966 & 0.245 & 0.254 \\
  $6-7$   & 4.74 & 0.0273 & 48.0 & 0.967 & 0.249 & 0.257 \\
  $7-8$   & 4.97 & 0.0284 & 52.3 & 0.961 & 0.228 & 0.237 \\
  $8-9$   & 5.20 & 0.0293 & 56.5 & 0.955 & 0.211 & 0.211 \\
  $9-10$  & 5.45 & 0.0342 & 69.1 & 0.933 & 0.173 & 0.185 \\
  $10-11$ & 5.58 & 0.0368 & 76.1 & 0.920 & 0.157 & 0.171 \\
  $11-12$ & 5.52 & 0.0329 & 67.3 & 0.937 & 0.177 & 0.189 \\
  $12-13$ & 5.14 & 0.0242 & 46.1 & 0.970 & 0.259 & 0.267 \\
  $13-14$ & 4.77 & 0.0185 & 32.7 & 0.985 & 0.365 & 0.371 \\
  $14-15$ & 4.41 & 0.0149 & 24.4 & 0.991 & 0.489 & 0.493 \\
  $15-16$ & 4.25 & 0.0132 & 20.8 & 0.994 & 0.574 & 0.577 \\
  $16-17$ & 3.94 & 0.0124 & 18.1 & 0.995 & 0.660 & 0.633 \\
  \hline
  \end{tabular}
\end{center}
\tablefoot{$\Delta R$: radial range,
$B_\mathrm{turb}$: strength of the isotropic turbulent equipartition field,
$\langle n_\mathrm{e} \rangle$: average thermal density,
$\sigma_\mathrm{RM}$: Faraday dispersion,
$\mathrm{DP}_6$: predicted depolarization factor at \wave{6.2},
$\mathrm{DP}_{20}$: predicted depolarization factor at \wave{20.5},
$\mathrm{DP}=\mathrm{DP}_{20}/\mathrm{DP}_6$,
all at $3\arcmin$ resolution.
}
\end{table}

\end{appendix}

\end{document}